\documentclass[a4paper,10pt]{article}
\pdfoutput=1 

\usepackage{jheppub} 
\usepackage{graphicx,floatflt,amssymb,rotate,epsfig}
\usepackage[utf8]{inputenc}
\usepackage[inline]{enumitem}
\usepackage{mathrsfs}
\usepackage{amssymb}
\usepackage{amsmath}
\usepackage{color}
\usepackage{graphicx}
\usepackage{subfig}
\usepackage{multirow}
\usepackage{mathtools}
\usepackage{float}
\usepackage{pstricks}
\usepackage[toc,page]{appendix}
\usepackage{pifont}
\usepackage{braket}
\usepackage{bbold}
\usepackage{hyperref}
\usepackage{newfloat}
\usepackage{pdflscape}
\DeclareCaptionType{equ}[][]

\usepackage{relsize}
\def\BaBar{{\mbox{\slshape B\kern-0.1em{\smaller A}\kern-0.1em B\kern-0.1em{\smaller A\kern-0.2em R}}}}

\newcommand{\ba}{\begin{array}}
	\newcommand{\ea}{\end{array}}
\def\beq{\begin{equation}}
	\def\eeq{\end{equation}}
\def\bea{\begin{eqnarray}}
	\def\eea{\end{eqnarray}}
\def\nn{\nonumber}

\def\roughly#1{\mathrel{\raise.3ex\hbox
		{$#1$\kern-.75em\lower1ex\hbox{$\sim$}}}}
\def\lsim{\roughly<}
\def\gsim{\roughly>}
\def\sla#1{\raise.15ex\hbox{$/$}\kern-.57em #1}

\def\bd{B_d^0}

\def\order{\lower 1.8ex \hbox{\LARGE\~{}}}

\def\btopi0lnu{B^+ \to \pi^0 \ell\nu}

\def\btorho0nu{B^+ \to \rho^0 \ell\nu}

\newcommand*{\rom}[1]{\expandafter\@slowromancap\romannumeral #1@}

\def\bd0tau{B\to D \tau\nu_{\tau}}

\def\be {\begin{equation}}
	\def\ee {\end{equation}}

\usepackage[normalem]{ulem}

\definecolor{darkgreen}{cmyk}{1,0,1,0.4}

\definecolor{pink}{cmyk}{0.4,1,0.3,0}

\def\com2#1{\textcolor{red}{\it{#1}}}

\title{\boldmath Extractions of $|V_{ub}|/|V_{cb}|$ from a combined study of the exclusive $b\to u(c) \ell ^- \bar{\nu}_{\ell}$ decays }

\author[a,b]{Aritra Biswas,}
\emailAdd{iluvnpur@gmail.com}
\author[a]{Soumitra Nandi,}
\emailAdd{soumitra.nandi@iitg.ac.in}
\author[a,c]{and Ipsita Ray}
\emailAdd{ipsitaray02@gmail.com}

\affiliation[a]{Department of Physics, Indian Institute of Technology Guwahati, Assam 781039, India}
\affiliation[b]{ Institut de Física d'Altes Energies (IFAE), Campus UAB, Facultat Ciencies Nord, 08193 Bellaterra, Barcelona}
\affiliation[c]{Department of Physics, Indian Institute of Technology Gandhinagar, Gujarat 382355, India}

\abstract{ We have extracted the ratio $|V_{ub}|/|V_{cb}|$ from a combined study of the available inputs on $\bar{B}\to (\pi,\rho,\omega)\ell^- \bar{\nu}_{\ell}$ and $\bar{B}(\bar{B}_s)\to (D^{(\ast)}, D_s^{(\ast)})\ell^- \bar{\nu}_{\ell}$ ($\ell = \mu$ or $e$) decays. We have done our analysis with and without the inclusion of the available experimental results on the branching fraction $\mathcal{BR}(\bar{B}_s\to K^+\mu^-\bar{\nu}_{\mu})$ or the ratio of the rates $\Gamma(\bar{B}_s\to K^+ \ell^-\bar{\nu}_{\ell})/ \Gamma(\bar{B}_s \to D_s^+ \ell^- \bar{\nu}_{\ell})$. We have used all the available experimental data on the branching fractions in different $q^2$-bins, the available theory inputs on the form factors from lattice and the light cone sum rule (LCSR) approach; analysed the data in different possible scenarios and presented the results. From a combined analysis of all the $b\to c\ell^- \bar{\nu}_{\ell}$ exclusive decays, we have obtained $|V_{cb}|=(40.5 \pm  0.6)\times 10^{-3}$ and from  all the $b\to u\ell^- \bar{\nu}_{\ell}$ modes, we obtain $|V_{ub}|=(3.46\pm 0.12)\times 10^{-3}$ which changes to $|V_{ub}|=(3.51 \pm 0.13)\times 10^{-3}$ after dropping $\mathcal{BR}(\bar{B}_s\to K^+\mu^-\bar{\nu}_{\mu})$. We hence report the values of the ratio $|V_{ub}|/|V_{cb}| = 0.085 \pm 0.003$ and $0.087 \pm 0.003$ with and without the input from $\bar{B}_s\to K^+\mu^-\bar{\nu}_{\mu}$ modes, respectively. We have also predicted the ratio $\Gamma(\bar{B}_s\to K^+ \ell^-\bar{\nu}_{\ell})/ \Gamma(\bar{B}_s \to D_s^+ \ell^- \bar{\nu}_{\ell})$ using the fit results. In addition, we provide the predictions of $\mathcal{BR}(\bar{B}_s\to K^+\mu^-\bar{\nu}_{\mu})$ and $\mathcal{BR}(\bar{B}\to (\rho,\omega)\ell^- \bar{\nu}_{\ell})$ in the Standard Model (SM) in small $q^2$-bins. }
   
\arxivnumber{}

\begin{document}
 \maketitle
	
\section{Introduction}

The Cabibbo-Kobayashi-Maskawa (CKM) matrix plays a central role in the understanding of the mechanism of quark mixing within the Standard Model. Hence precise determination of its elements is essential. An accurate extraction of these elements have received special attention in both theoretical and experimental capacities in the last few decades. The unitarity relation $\sum_{k} V_{ik}^*V_{kj} = 0$ leads to six unitarity triangles; the area of each being a measure of CP-violation. One of these triangles is obtained from the relation: 
\begin{equation}\label{eq:UT1}
 V_{ub}^*V_{ud} + V_{cb}^*V_{cd} + V_{tb}^*V_{td}  = 0.
\end{equation}  
Using the above relation one can obtain a triangle whose sides are of equal magnitudes, and one of the primary goals of the flavour community is the precise construction of this triangle \cite{HFLAV:2022pwe}. To date, the elements $|V_{ub}|$ and $|V_{cb}|$ are known to a relatively less degree of precision as compared to the other CKM elements required to construct the above triangle. The decay rates of the semileptonic $b\to c\ell^- \bar{\nu}_{\ell}$ and $b\to u\ell^- \bar{\nu}_{\ell}$ transitions are sensitive to $|V_{cb}|^2$ and $|V_{ub}|^2$ respectively, and are hence useful for the extractions of these CKM elements. In this regard, the corresponding exclusive and inclusive decays play essential roles. Concerning the exclusive decays, we have a good amount of information available on the $\bar{B}\to \pi\ell^- \bar{\nu}_{\ell}$, $\bar{B}\to D^{(*)}\ell^- \bar{\nu}_{\ell}$ and $\bar{B}_s\to D_s^{(*)}\ell^- \bar{\nu}_{\ell}$ \footnote{Charge-conjugate decays are implied.} decays \cite{pdg}. So far, $|V_{ub}|$ has been extracted using the available information on $\bar{B}\to \pi\ell^- \bar{\nu}_{\ell}$ decays \cite{HFLAV:2022pwe,pdg,Biswas:2020uaq,Martinelli:2022tte} while $|V_{cb}|$ has been extracted independently from the $\bar{B}\to D^{(*)}\ell^- \bar{\nu}_{\ell}$ \cite{Belle:2015pkj,MILC:2015uhg,Na:2015kha,Abdesselam:2017kjf,Belle:2018ezy,FermilabLattice:2021cdg} and $\bar{B}_s\to D_s^{(*)}\ell^- \bar{\nu}_{\ell}$ \cite{LHCb:2020cyw} decays. The inclusive determination of $|V_{ub}|$ is heavily dependent on the QCD modelling of the non-perturbative shape function; for details see the refs. \cite{BLNP2005,DGE2005,GGOU2007,ADFR}. On the contrary, the inclusive determination of $|V_{cb}|$ is relatively clean, see for example the refs.~\cite{Alberti:2014yda,Gambino:2016jkc,Bordone:2021oof,Bernlochner:2022ucr}. As mentioned above, the up-to-date values of these elements extracted from exclusive and inclusive decays can be seen from HFLAV or PDG \cite{HFLAV:2022pwe,pdg} and the references therein. It has been observed that the inclusive determinations of these two elements have higher values concerning the corresponding values extracted from exclusive decays. The difference is higher in the extracted values of $|V_{ub}|$. No convincing explanations for the scenario are available till date.

For a better understanding of the observed discrepancies, it is important to extract the CKM elements from other modes. For example, $|V_{cb}|$ can be extracted separately from $\bar{B}_s\to D_s^{(*)}\ell^- \bar{\nu}_{\ell}$ and $\Lambda_b\to \Lambda_c^+\ell^- \bar{\nu}_{\ell}$ decay modes apart from the $\bar{B}\to D^{(*)}\ell^- \bar{\nu}_{\ell}$ decays and the extracted values can be compared. The absence of light spectator quarks in $\bar{B}_s\to D_s^{(*)}\ell^- \bar{\nu}_{\ell}$ transitions makes them more attractive from the theoretical viewpoint as the statistical errors in Lattice QCD computations, and finite volume effects get highly reduced. After analysing the available data on $\bar{B}_s\to D_s^{(*)}\ell^- \bar{\nu}_{\ell}$ decay, the LHCb collaboration had obtained $|V_{cb}| = (42.3 \pm 1.4)\times 10^{-3}$ \cite{LHCb:2020cyw} which is consistent with that extracted from $\bar{B}\to D^{(*)}\ell^- \bar{\nu}_{\ell}$ decays \cite{HFLAV:2022pwe}. Similarly, apart from the $\bar{B} \to \pi l^- \bar{\nu}$ decay modes, the study of the additional modes like $\bar{B} \to \rho l^- \bar{\nu}$ and $\bar{B} \to \omega l^- \bar{\nu}$ would provide complementary information on the estimates of $|V_{ub}|$. Equally important are the measurements of the ratios of the leptonic or semileptonic decay widths sensitive to the ratio of $|V_{ub}|$ and ${|V_{cb}|}$.  
For example, the ratio of the CKM elements $|V_{ub}|/{|V_{cb}|}$ has been measured by the LHCb collaboration from the measurements of the ratio of the decay width $\Gamma(\Lambda_b \to p \mu^- \bar{\nu})/\Gamma(\Lambda_b \to \Lambda_c \mu^- \bar{\nu})$ \cite{LHCb:2015eia,pdg}, and their measured value is given by  
\begin{equation}\label{eq:VubVcbbaryon}
|V_{ub}|/|V_{cb}| = 0.079 \pm 0.004_{\text{exp}} \pm 0.004_{\text{th}}. 
\end{equation}
Recently, there has also been a measurement of the ratio $R_{BF}=\mathcal{BR}(\bar{B}_s \to K^+ \mu^- \bar{\nu})/\mathcal{BR}(\bar{B}_s \to D_s^+ \mu^- \bar{\nu})$ from LHCb \cite{LHCb:2020ist} in two bins of the $B_s \to K^-$ momentum transfer, namely $q^2$ $<$ 7 $\text{GeV}^2$ and $q^2$ $>$ 7 $\text{GeV}^2$ with $B_s \to D_s \mu \nu$ acting as the normalization channel. The respective measured values are as given below
	\begin{eqnarray} \label{eq:ratio}
	   R_{BF}^{low} &=& (1.66 \pm 0.08(\text{stat}) \pm 0.07(\text{syst}) \pm 0.05(D_s)) \times 10^{-3}, \nonumber \\
	   R_{BF}^{\text{high}} &=& (3.25 \pm 0.21(\text{stat})^{+0.16}_ {-0.17}(\text{syst}) \pm 0.09(D_s)) \times 10^{-3}
	\end{eqnarray}
	where the uncertainties correspond to  statistical, systematic and due to the $D_s^- \to K^+ K^- \pi^-$ branching fraction. After incorporating the theory calculations on the form factors, the branching ratios as mentioned above are used to extract the ratio $|V_{ub}|/|V_{cb}|$ which are given as follows:
\begin{eqnarray} \label{eq:vubvcb}
	|V_{ub}|/|V_{cb}|^{low} &=& 0.061 \pm 0.002_{\text{exp}} \pm 0.003_{\text{th}}, \nonumber \\
	|V_{ub}|/|V_{cb}|^{high} &=& 0.095 \pm 0.004_{\text{exp}} \pm 0.007_{\text{th}}.  
\end{eqnarray}
Here, the theoretical uncertainties correspond to $D_s^-$ branching fraction and the form factor integrals, for details see \cite{LHCb:2020ist}. According to the LHCb collaboration, the difference in the predictions in the two bins are related to the differences in the theory inputs in the high (lattice) and low-bins (LCSR), respectively.  

The measurements of the ratio $|V_{ub}|/|V_{cb}|$ can be directly used to construct the unitarity triangle defined via eq.~\ref{eq:UT1}. However, these measurements could also be helpful to constrain $|V_{ub}|$ once we have the precise estimates of $|V_{cb}|$ and the above ratio or vice versa. This kind of exercise will provide complementary information concerning the different measurements and may help to understand the discrepancies between the inclusive and exclusive determinations.

In this paper, we have extracted $|V_{cb}|$ from a combined study of $\bar{B}\to D^{(*)}\ell^- \bar{\nu}_{\ell}$ and $\bar{B}_s\to D_s^{(*)}\ell^- \bar{\nu}_{\ell}$ decays after incorporating all the available experimental data \cite{Belle:2015pkj,Belle:2018ezy,LHCb:2020cyw} and the information on the form factors from the theory like  lattice QCD \cite{Bailey:2012rr,MILC:2015uhg,Na:2015kha,McLean:2019qcx,Harrison:2021tol,FermilabLattice:2021cdg} and LCSR \cite{Gubernari:2018wyi,Bordone:2019guc}. We have also extracted $|V_{ub}|$ from independent and combined studies of the $\bar{B}\to (\pi, \rho,\omega) \ell^- \bar{\nu}_{\ell}$ decay modes, including the available experimental data \cite{delAmoSanchez:2010af,Ha:2010rf,BaBar:2012dvs,BaBar:2013pls,Lees:2012vv,Sibidanov:2013rkk} and the theory inputs from lattice \cite{FermilabLattice:2015cdh,Flynn:2015mha,Colquhoun:2022atw} and LCSR \cite{Straub:2015ica,PhysRevD.101.074035,Gubernari:2018wyi,Leljak:2021vte}. Along with these $b\to u \ell^- \bar{\nu}_{\ell}$ modes, we have further analyzed the available information on $\bar{B}_s\to K^+ \mu^- \bar{\nu}_{\mu}$ decay, like the $q^2$ integrated branching fraction measured by LHCb \cite{LHCb:2020ist}
\begin{equation}\label{eq:totalBstok}
\mathcal{BR}(\bar{B}_s\to K^+ \mu^- \bar{\nu}_{\mu}) = (1.06 \pm 0.05_{\text{stat}} \pm 0.08_{\text{syst}})\times 10^{-4},
\end{equation}
and the inputs on the form factors from lattice \cite{PhysRevD.100.034501,HPQCD:BStoK2014,Flynn:2015mha} and LCSR \cite{Khodjamirian:2017fxg}. For a comparative study, we have defined a few distinct fit scenarios with and without the inputs from $\bar{B}_s\to K^+ \mu^- \bar{\nu}_{\mu}$ decay. In all these fits, we have extracted the ratio $|V_{ub}|/|V_{cb}|$ and predicted the ratio of the branching fractions: ${\mathcal{B}}(\bar{B}_s \to K^+ \mu^- \bar{\nu}_{\mu})/{\mathcal{B}}(\bar{B}_s \to D_s^+ \mu^- \bar{\nu}_{\mu})$ which can be compared to the respective measurements. The results obtained for $|V_{ub}|/|V_{cb}|$ from the semileptonic decays of $B$ and $B_s$ mesons have been compared with the one measured by LHCb from a measurement of the ratio of baryonic decays like $\Gamma(\Lambda_b \to p \mu^- \bar{\nu})/\Gamma(\Lambda_b \to \Lambda_c \mu^- \bar{\nu})$. Refs.~\cite{Gonzalez-Solis:2021pyh,Martinelli:2022tte} had focused on the extractions of $|V_{ub}|$ from the simultaneous studies of the $\bar{B}\to \pi\ell^- \bar{\nu}_{\ell}$ and $\bar{B}_s\to K \mu^- \bar{\nu}_{\mu}$ modes, we have a slightly different approach, and the results are in good agreement within the error bars.    

In some additional fits, alongside the available decay modes mentioned above, we have included the measured values of the ratio of branching fractions given in eq.~\ref{eq:ratio} and the theory inputs on the form factors for $\bar{B}_s \to K$ decays from lattice \cite{PhysRevD.100.034501,HPQCD:BStoK2014,Flynn:2015mha} and LCSR \cite{Khodjamirian:2017fxg}. In these fits, we have not included the measured value of ${\mathcal{B}}(\bar{B}_s \to K \mu^- \bar{\nu}_{\mu})$ as input since the measured values for the ratios are obtained from the measurement of $\mathcal{B}(\bar{B}_s\to K \mu^- \bar{\nu}_{\mu})$ after normalizing it by the measured branching fraction $\mathcal{B}(\bar{B}_s\to D_s \mu^- \bar{\nu}_{\mu})$. Furthermore, in these fits, we have extracted $|V_{ub}|$, $|V_{cb}|$ and the correlation between them, along with their correlations with all the form factor parameters. Including all these pieces of information, we subsequently estimated the ratio $R_{BF}$ (for a check) and the $|V_{ub}|/|V_{cb}|$.

\section{Theory}

\subsection{B decays to pseudoscalar mesons}
 
\begin{table}[ht]
	\begin{center}
		\begin{tabular}{|c|c|}
			\hline
			Constants & Values\\
			\hline
			$G_F$ & 1.166 $\times$ $10^{-5}$ ~GeV$^{-2}$\\
			$\chi^{T}_{1^-}$ (0)  (for $g$ and $f_+$) & 5.84 $\times$ $10^{-4}$~GeV$^{-2}$ \\
			$\chi^{T}_{1^+}$ (0)  (for $f$ and $F_1$) & 4.69 $\times$ $10^{-4}$~GeV$^{-2}$ \\
			$\chi^{L}_{0^-}$ (0)  (for $F_2$) & 2.19 $\times$ $10^{-2}$~GeV$^{-2}$ \\
			$\chi^{L}_{0^+}$ (0)  (for $f_0$) & 7.58 $\times$ $10^{-3}$~GeV$^{-2}$ \\
			\hline
		\end{tabular}
		\caption{Various inputs used in this analysis \cite{Martinelli:2022xir}. }
		\label{tab:inputs}
	\end{center}
\end{table} 

\begin{table}[ht]
	\begin{center}
		\begin{tabular}{|c|c|}
			\hline
			Form factor involved & $B_c^{(*)}$ pole masses (GeV)\\
			\hline
			$f_+$ and $g$ & 6.32847, 6.91947, 7.030\\
			$f$ and $F_1$ &  6.73847, 6.750, 7.145, 7.150\\
			$F_2$ & 6.27447, 6.8712, 7.250\\
			$f_0$ & 6.70347, 7.122\\
			\hline
		\end{tabular}
		\caption{Pole masses used in the $\bar{B} \to D^{(*)}$ modes.}
		\label{tab:poleBD}
	\end{center}
\end{table} 

\begin{table}[htbp]
	\begin{center}
		\begin{tabular}{|c|c|}
			\hline
			Form factor involved & $B_c^{(*)}$ pole masses (GeV)\\
			\hline
			$f_+$  & 6.32847, 6.91947, 7.030, 7.280\\
			$g$ & 6.32847, 6.91947, 7.030, 7.280\\
			$f$ and $F_1$ &  6.73847, 6.750, 7.145, 7.150\\
			$F_2$ & 6.27447, 6.8712, 7.250\\
			$f_0$ & 6.70347, 7.122\\
			\hline
		\end{tabular}
		\caption{Pole masses used in the $\bar{B}_s \to D_s^{(*)}$ modes.}
		\label{tab:poleBs}
	\end{center}
\end{table} 
As mentioned in the introduction, we have considered the semileptonic decays of $B$ and $B_s$ mesons to pseudoscalar and vector mesons in this analysis. The expressions for the corresponding decay rate distributions are given in refs.~\cite{Boyd:1997kz,Belle:2018ezy,Becirevic:2019tpx,Alguero:2020ukk,Biswas:2021pic}. The respective rates of the $B$ and $B_s$ decays to pseudoscalar mesons are dependent on the form factors $f_{+,0}(q^2)$ \cite{Sakaki:2013bfa} while those for the decays to vector mesons depend on the four form factors $A_{0,1,2}(q^2)$ and $V(q^2)$. These form factors are non-perturbative quantities and the major sources of theoretical uncertainties in the decay rates. The non-perturbative methods like lattice QCD and LCSR are useful to predict the shape of the form factors within the kinematically allowed regions. Note that lattice QCD calculations predict the form factors at high-$q^2$ values while the calculations based on the LCSR approach are more reliable near $q^2=0$. However, we need to determine the shape of the form factors in the whole kinematically allowed $q^2$ region. Therefore, to determine the shape of the $q^2$ spectrum of the form factors in heavy-to-heavy transitions like $\bar{B} \to D^{(*)}$ and $\bar{B}_s \to D_s^{(*)}$ in physically allowed regions, we use the following $z(q^2)$-expansion as was suggested in \cite{Boyd:1997kz} by Boyd-Grinstein-Lebed (BGL):
\begin{equation}
f_i (q^2) = \frac{1}{P_i (q^2) \phi_i (q^2,t_0)} \sum_{j=0}^{N} a_{j}^i z^j. 
\label{eq:FF-BGL}
\end{equation}
 The conformal map from $q^2$ to $z \equiv z[q^2; t_0]$ is given by :
 \begin{equation}\label{eq:z}
 z(q^2,t_0) = \frac{\sqrt{t_+-q^2}-\sqrt{t_+-t_0}}{\sqrt{t_+-q^2}+\sqrt{t_+-t_0}}\,,  
 \end{equation}
 where
 $t_\pm \equiv (m_B\pm m_F)^2$ and $t_0\equiv t_+(1-\sqrt{1-t_-/t_+})$. $F$ refers to the final state meson. $t_0$ is a free parameter that governs the size of $z$ in the semileptonic phase space. This parameterization of the form factors relies on a Taylor series expansion about $z=0$. The key ingredient in this approach is the transformation that maps the complex $q^2$ plane onto the unit disc $|z| \le 1$. The coefficients $a_j^{i}$ follow weak as well as strong unitarity constraints \cite{Boyd:1997kz}. Here, $f_i(q^2)$ includes the form-factors $f^{(s)}_+(q^2)$ and $f^{(s)}_0(q^2)$ relevant for the $\bar{B} (\bar{B}_s) \to D (D_s)$ decays, along with the form factors $F^{(s)}_1(q^2)$, $f^{(s)}(q^2)$, $g^{(s)}(q^2)$ and $F^{(s)}_2(q^2)$ associated to the $\bar{B} (\bar{B}_s) \to D^* (D^*_s)$ decays. Note that $F_1$, $f$, $g$ and $F_2$ are related to $A_{0,1,2}$ and $V$ which can be verified, for example, from ref.~\cite{Gambino:2020jvv}.  

In eq.~\ref{eq:FF-BGL}, the variable $z$ is also related to the recoil variable $w$ ($w$ = $v_B.v_F$ with $v_B$ and $v_F$ being the four-velocities of the $B$ and the final state mesons respectively) as 
\begin{equation}
z = \frac{\sqrt{w+1}-\sqrt{2}}{\sqrt{w+1}+\sqrt{2}},
\end{equation}
where $w$ is related to the momentum transferred to the dilepton system ($q^2$) as $q^2 = m_B^2 + m_F^2 - 2 m_B m_F w$. The Blaschke factor $P_i (q^2)$ accounts for the poles in $f_i(q^2)$ for the subthreshold resonances with the same spin-parity as the current defining the form factors. Mathematically, the $P_i (q^2)$ is defined as    
\begin{equation}
P_i(q^2) = \prod_p  z[q^2;t_p] ,
\label{eq:Blaschke-fact}
\end{equation}
where for a single pole
\begin{equation}
z[q^2;t_p] = \frac{z-z_p}{1 - z z_p}, \text{with}\ \  z_p \equiv z[t_p; t_0].
\end{equation} 
The role of the $P_i (q^2)$ is to eliminate the poles in $f_i(q^2)$ at the resonance masses below the threshold. Following eq.~\ref{eq:z}, we will get 
\begin{equation}
z_p = \frac{\sqrt{(m_B + m_F)^2 - m_P^2} - \sqrt{4 m_B m_F}}{\sqrt{(m_B + m_F)^2 - m_P^2} + \sqrt{4 m_B m_F}}.
\label{eq:zp}
\end{equation}
Here, $m_P$ denotes the pole masses. The detailed calculation of the outer functions $\phi_i (q^2;t_0)$ are available in \cite{Boyd:1997kz}. The numerical inputs for the perturbatively calculable $\chi$-functions along with other relevant inputs for the outer functions are taken from \cite{Martinelli:2022xir,Boyd:1997kz,Bigi:2017jbd,Jaiswal:2020wer}. In table \ref{tab:inputs}, we give the values of the $\chi$ functions relevant to the different form factors. The pole masses used in $\bar{B} \to D^{(*)}$ and $\bar{B}_s \to D_s^{(*)}$ channels are given in tables \ref{tab:poleBD} and \ref{tab:poleBs}, respectively.

 To determine the shape of the form factors for $\bar{B} \to (\pi,\rho,\omega)\ell^- \bar{\nu}_{\ell}$  and $\bar{B}_s\to K \ell^- \bar{\nu}_{\ell}$ decays, we use the simplified series expansion in $z$ as given in ref.~\cite{Straub:2015ica} by Bharucha-Straub-Zwicky (BSZ): 
\begin{equation}\label{eq:bszexp}
 f_i(q^2) = \frac{1}{1 - q^2/m_{R,i}^2} \sum_{k=0}^N a_k^i \, [z(q^2;t_0)-z(0;t_0)]^k. 
 \end{equation}
In eq \ref{eq:bszexp},  $m_{R,i}$ denotes the mass of the sub-threshold resonances compatible with the quantum numbers of the respective form factors and $a_k^i$s are the coefficients of expansion, the details are given in ref \cite{Straub:2015ica}. Another commonly used approach for the parametrizations of the $\bar{B}\to \pi$ and $\bar{B}_s\to K$ form factors in the literature is provided by Bourrely-Caprini-Lellouch (BCL)~\cite{BCL}. A comparative study of these two approaches in the extractions of $|V_{ub}|$ from $\bar{B}\to \pi\ell^- \bar{\nu}_{\ell}$ decays has been undertaken in ref.~\cite{Biswas:2021qyq}. The results are extremely consistent with each other in both approaches. As we have mentioned earlier, for $\bar{B}\to \rho$ and $\bar{B}\to \omega$ form factors only the inputs from LCSR approaches are available \cite{Straub:2015ica,Gubernari:2018wyi}. In these approaches, the pseudo data points are obtained following the parametrization given in eq. \ref{eq:bszexp}. We have used the simplified expansion as given above for all the heavy-to-light transitions in order to be consistent.

Note that there have been attempts to extract $|V_{ub}|$ from $\bar{B}\to \pi\ell^- \bar{\nu}_{\ell}$ decays using the BGL parametrization for the corresponding form factors \cite{Arnesen:2005ez}. However, as argued in ref.~\cite{Becher:2005bg,BCL} the dispersive bounds allow the higher order terms in the BGL expansion to contribute significantly (here $|z_{max}| \approx 0.3$) for heavy-to-light transitions. Therefore, one may need to consider more terms in the series, and the fit may not be stable with the currently available inputs. However, for the heavy-to-heavy transitions, the dispersive bounds suggest that only the first few terms in the series are required for a reasonable accuracy, which has already been observed in the fits to data in $\bar{B}\to D^{(*)}\ell^- \bar{\nu}_{\ell}$ decays \cite{Jaiswal:2017rve,Jaiswal:2020wer}.

\begin{table} [t]
	\centering
	\begin{tabular}{|c|c|c|c|}
		\hline
		\multicolumn{2}{|c|}{$N_{sig}$ for $B_s \to D_s \mu \nu$} & \multicolumn{2}{c|}{ $N_{sig}$ for $B_s \to D_s^* \mu \nu$}   \\
		\hline
		Bin & Value ($10^{-8}$) & Bin & Value ($10^{-8}$)\\
		\hline
		1 - 1.1 &  0.123 $\pm$ 0.011 & 1 - 1.1 & 1.962 $\pm$ 0.134\\
		1.1 - 1.2 & 0.496 $\pm$ 0.039 & 1.1 - 1.2 & 2.839 $\pm$ 0.267\\
		1.2 - 1.3 & 0.905 $\pm$ 0.064 & 1.2 - 1.3 & 2.854 $\pm$ 0.267\\
		1.3 - 1.4 & 1.278 $\pm$ 0.082 & 1.3 - 1.4 & 2.504 $\pm$ 0.236\\
		1.4 - 1.5 & 1.595 $\pm$ 0.095  & 1.4 - 1.46723 & 1.339 $\pm$ 0.161\\
		1.5 - 1.54667 & 0.835 $\pm$ 0.049 & &\\
		\hline
	\end{tabular}
	\caption{Synthetic data created for $\bar{B}_s \to D_s^{(*)}$ modes \cite{LHCb:2020cyw}.} 
	\label{tab:syntheticlhcb}
\end{table}

\begin{table}[t]
	\begin{center}
		\small
		\renewcommand*{\arraystretch}{1.6}
		\begin{tabular}{|c|c|c|}
			\hline
			Fit parameters & Fit results from \cite{LHCb:2020cyw} & Our fit using table \ref{tab:syntheticlhcb} \\
			\hline
			$|V_{cb}| \times 10^{3}$ & 42.3 $\pm$ 1.4 & 42.3 $\pm$ 1.2\\
			$a_0$ & 0.037 $\pm$ 0.009 & 0.037 $\pm$ 0.012\\
			$a_1$ & 0.28 $\pm$ 0.27 & 0.29 $\pm$ 0.32\\
			$b_1$ & -0.06 $\pm$ 0.07 & -0.06 $\pm$ 0.10\\
			$c_1$ & 0.0031 $\pm$ 0.0023 & 0.0027 $\pm$ 0.0030\\
			$\mathcal{G}(0)$ & 1.094 $\pm$ 0.034 & 1.081 $\pm$ 0.031 \\
			$d_1$ &  -0.017 $\pm$ 0.008 & -0.014 $\pm$ 0.006 \\
			$d_2$ & -0.26 $\pm$ 0.05 & -0.25 $\pm$ 0.04\\
			$h_{A_1}(1)$ & 0.900 $\pm$ 0.013 & 0.902 $\pm$ 0.012\\
			\hline
		\end{tabular}
		\caption{Comparison of the fit results given in ref \cite{LHCb:2020cyw} with those obtained from a fit to the synthetic data generated in this analysis using the results of \cite{LHCb:2020cyw}. }
		\label{tab:lhcbfitresult}
	\end{center}
\end{table} 

\section{Analysis and Results}

In this section, we discuss the details of our analyses along with the individual results. In the first few analyses, we independently extract $|V_{cb}|$ or $|V_{ub}|$ from the respective modes, as mentioned earlier. Following these, we perform an extraction from the different combined analyses. We also carry out separate analyses with the different inputs for the form factors from the different lattice groups and LCSR, which help us understand the impact of different theory inputs. For all of these individual fits, we predict the ratio $|V_{ub}|/|V_{cb}|$ and the ratio of branching fractions: 
\begin{equation}
R_{BF} =\frac{\text{BR}(\bar{B}_s \to K^+ \mu^- \bar{\nu})}{\text{BR}(\bar{B}_s \to D_s^+ \mu^- \bar{\nu})}, 
\end{equation}
in the low and high $q^2$ bins and compare with the measured values given in eq.~\ref{eq:ratio}. Finally, we carry out combined fits for all the available inputs on $b\to c\ell^- \bar{\nu}_{\ell}$ and $b\to u\ell^- \bar{\nu}_{\ell}$ decay modes with and without the measured values given in eq.~\ref{eq:ratio}, and predict $|V_{ub}|/|V_{cb}|$. All the fits and subsequent analyses have been carried out using a \emph{Mathematica\textsuperscript \textregistered} package~\cite{optex}. In the following subsections, we discuss the details of the available inputs, the different fit scenarios and the respective results for the different analyses.

\subsection{$|V_{cb}|$ from $\bar{B} \to D^{(*)} \ell^- \bar{\nu}_{\ell}$ and $\bar{B}_s \to D_s^{(*)} \ell^- \bar{\nu}_{\ell}$ decays }	

\begin{table} [t]
	\begin{center}
		\small
		\renewcommand*{\arraystretch}{1.2}
		\begin{tabular}{|c|c|c|c|c|}
			\hline
			Mode & Inputs & $\chi^2_{min}$/DOF & p-value (\%) & $|V_{cb}|$ $\times$ $10^3$\\
			\hline
			$\bar{B} \to D l^- \bar{\nu}$ & Experiment+Fermilab/MILC \cite{MILC:2015uhg} +  & 33.0/47 & 93.9 & 41.4 $\pm$ 1.1\\
			& HPQCD \cite{Na:2015kha}  + LCSR \cite{Gubernari:2018wyi} & & & \\
			
			$\bar{B} \to D^* l^- \bar{\nu}$ &  Experiment+Fermilab-MILC \cite{FermilabLattice:2021cdg} + & 120.5/81 & 0.3  & 39.2 $\pm$ 0.8 \\
			
			& LCSR \cite{Gubernari:2018wyi} & & & \\
			\hline
			$\bar{B}_s \to D_s^{(*)} l^- \bar{\nu}$ & Experiment+ HPQCD \cite{McLean:2019qcx,Harrison:2021tol} + & 2.98/17 & 99.9  & 42.2 $\pm$ 1.3 \\
			& LCSR \cite{Bordone:2019guc} & & &\\
			
			& Experiment+ Fermilab/MILC \cite{Bailey:2012rr} + & 29.3/20 & 8.2 & 44.0 $\pm$ 1.3 \\
			& HPQCD \cite{McLean:2019qcx,Harrison:2021tol} +LCSR \cite{Bordone:2019guc} & & & \\
			
			\hline
			All modes &  All Experiments + All lattice + LCSR  & 273.9/156 & 1.7 $\times$ $10^{-6}$  & {\bf 40.5 $\pm$ 0.6}\\
			\cline{2-5}
			&  All Experiments + All lattice & 163.2/150 & 21.8 & 40.3 $\pm$ 0.6 \\
			& (except $f_{+,0}^{B_s\to D_s}$ Fermilab/MILC \cite{Bailey:2012rr}) & & & \\
			&  +LCSR & & & \\
			\hline
		\end{tabular}
		\caption{A comparative study of the extraction of $|V_{cb}|$ in different channels. }
		\label{tab:Vcb}
	\end{center}
\end{table} 

\begin{figure*}[t]
	\small
	\centering
	\subfloat[]{\includegraphics[width=0.3\textwidth]{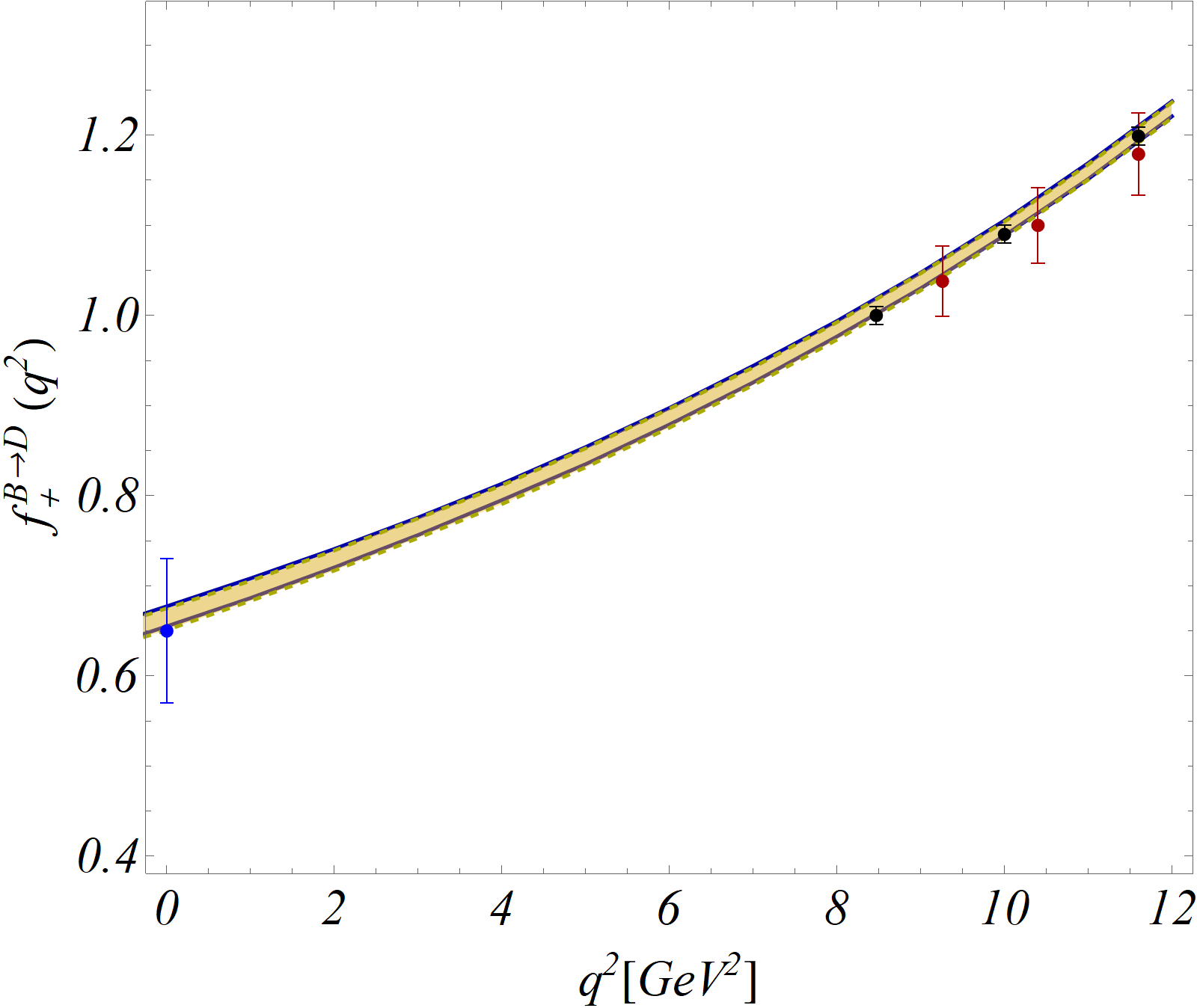}\label{fig:fplusB2D}}~~~
	\subfloat[]{\includegraphics[width=0.3\textwidth]{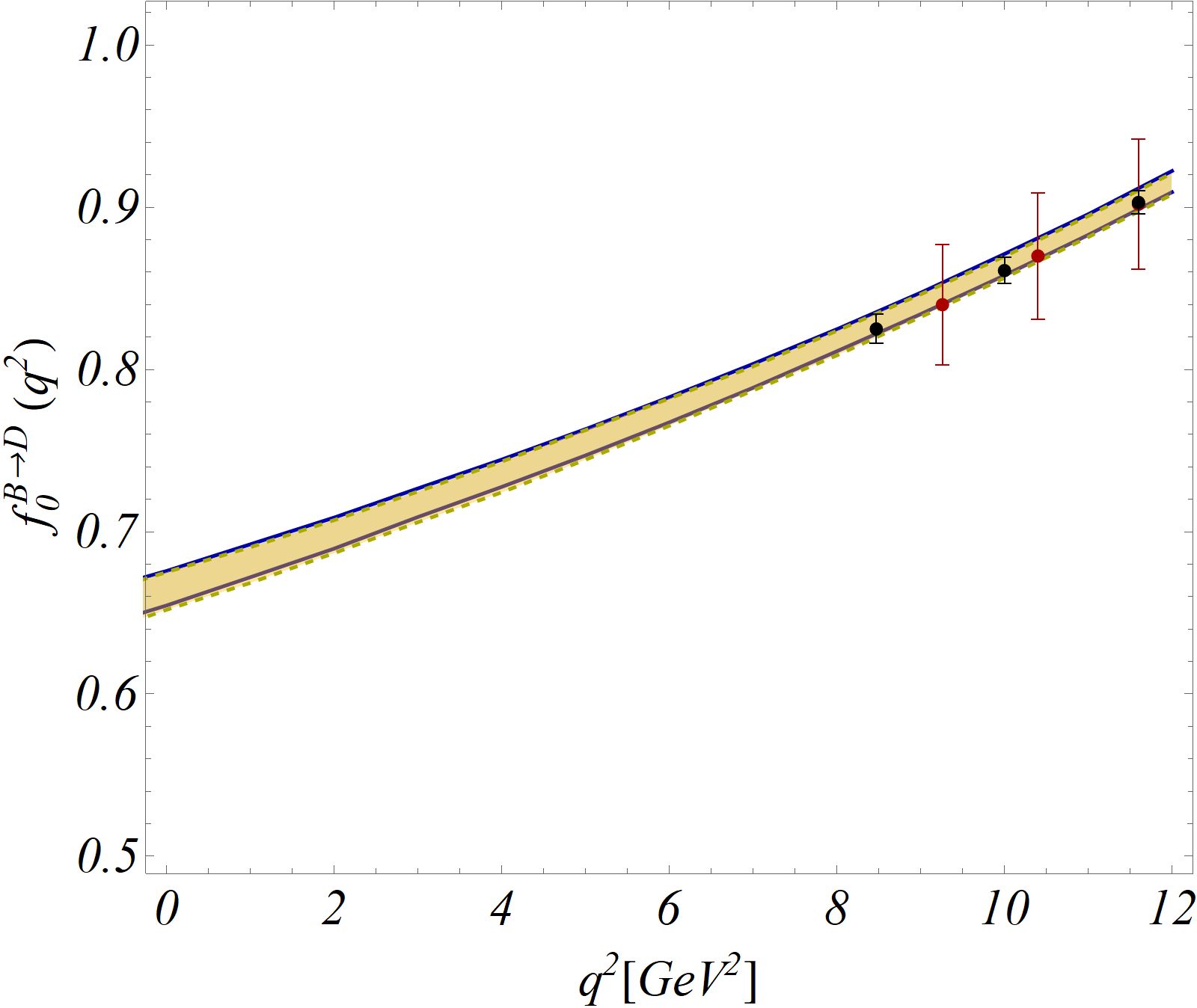}\label{fig:f0B2D}}~~~
	\subfloat[]{\includegraphics[width=0.3\textwidth]{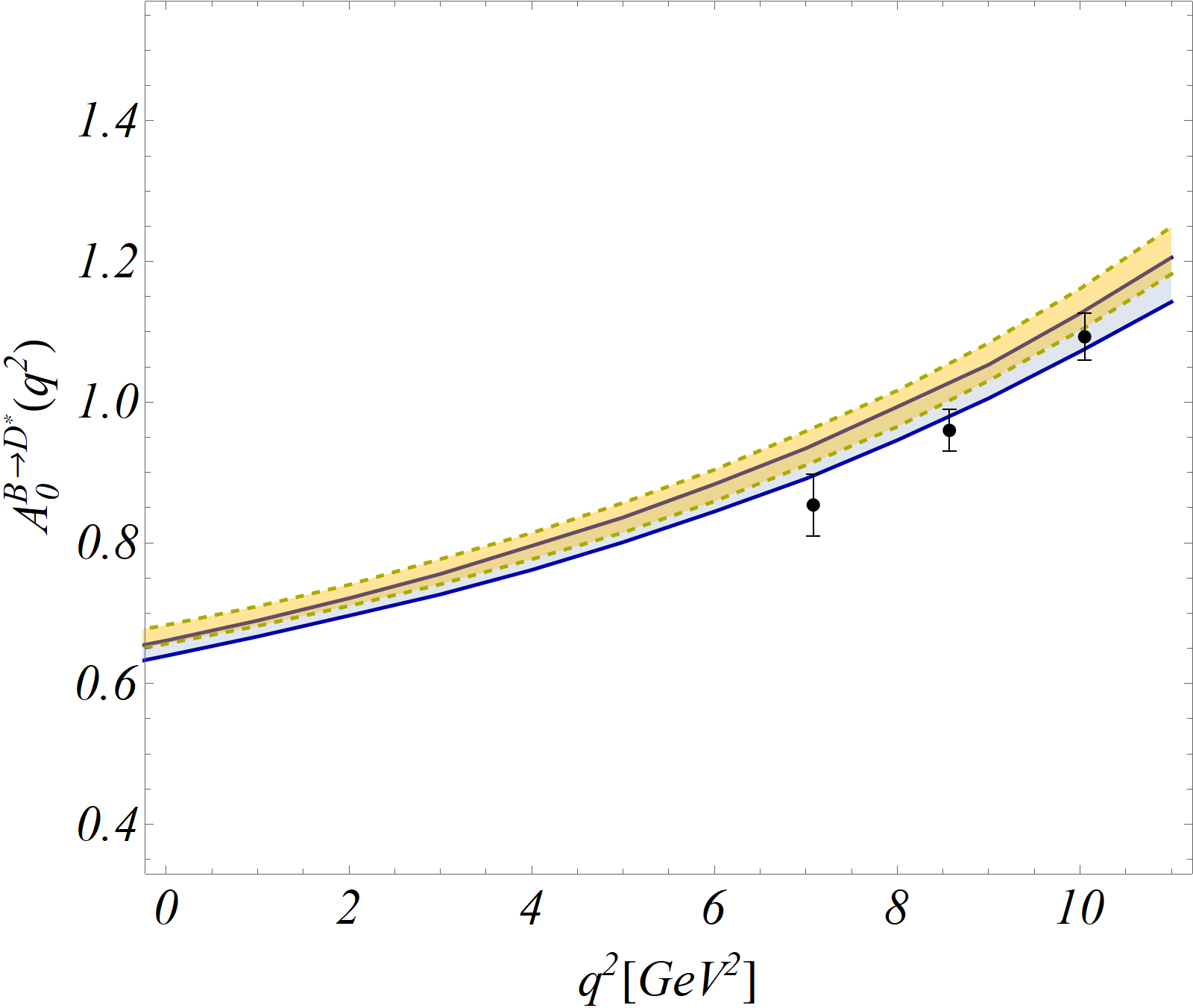}\label{fig:A0Dst}}\\
	\subfloat[]{\includegraphics[width=0.3\textwidth]{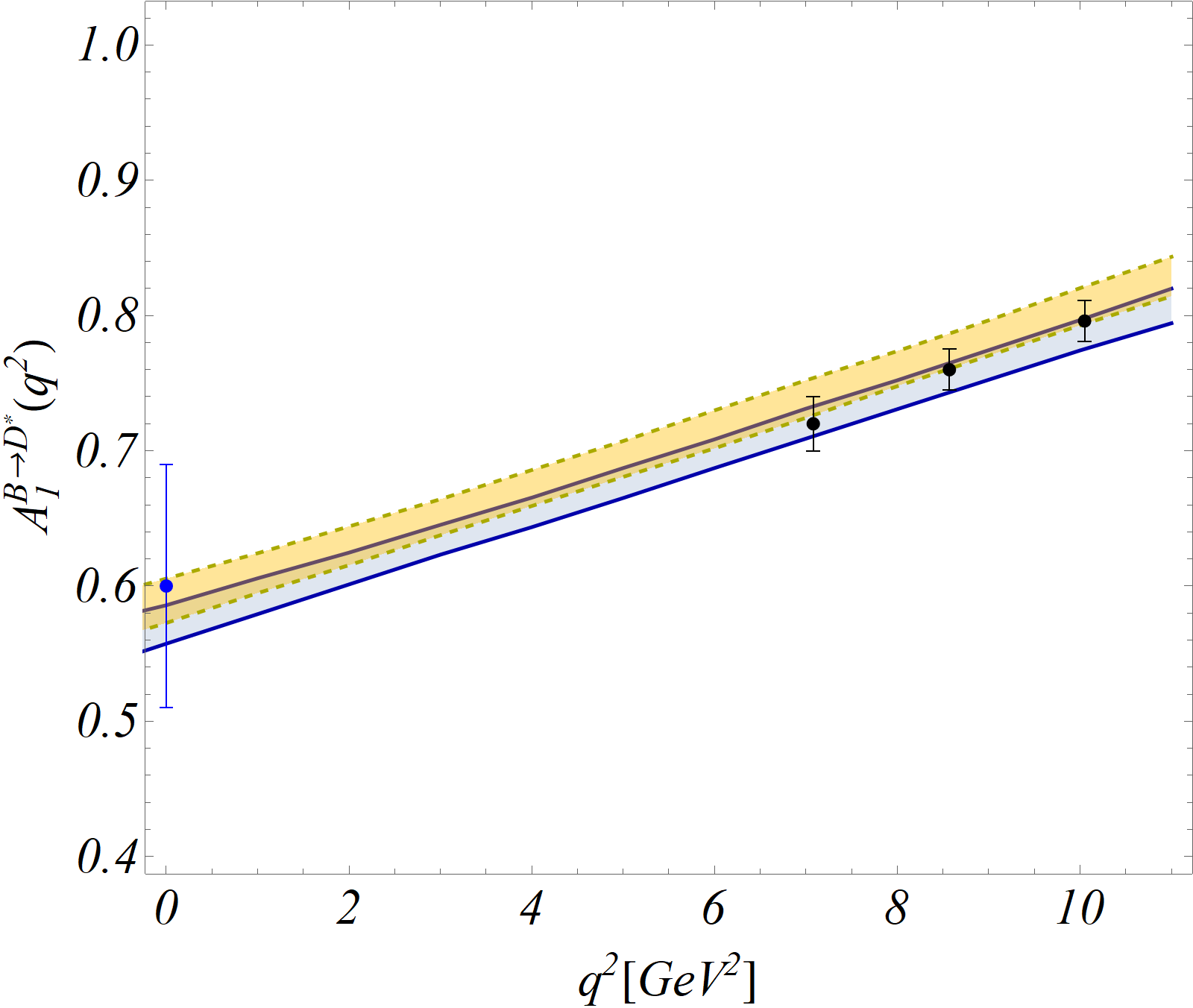}\label{fig:A1Dst}}~~~	\subfloat[]{\includegraphics[width=0.3\textwidth]{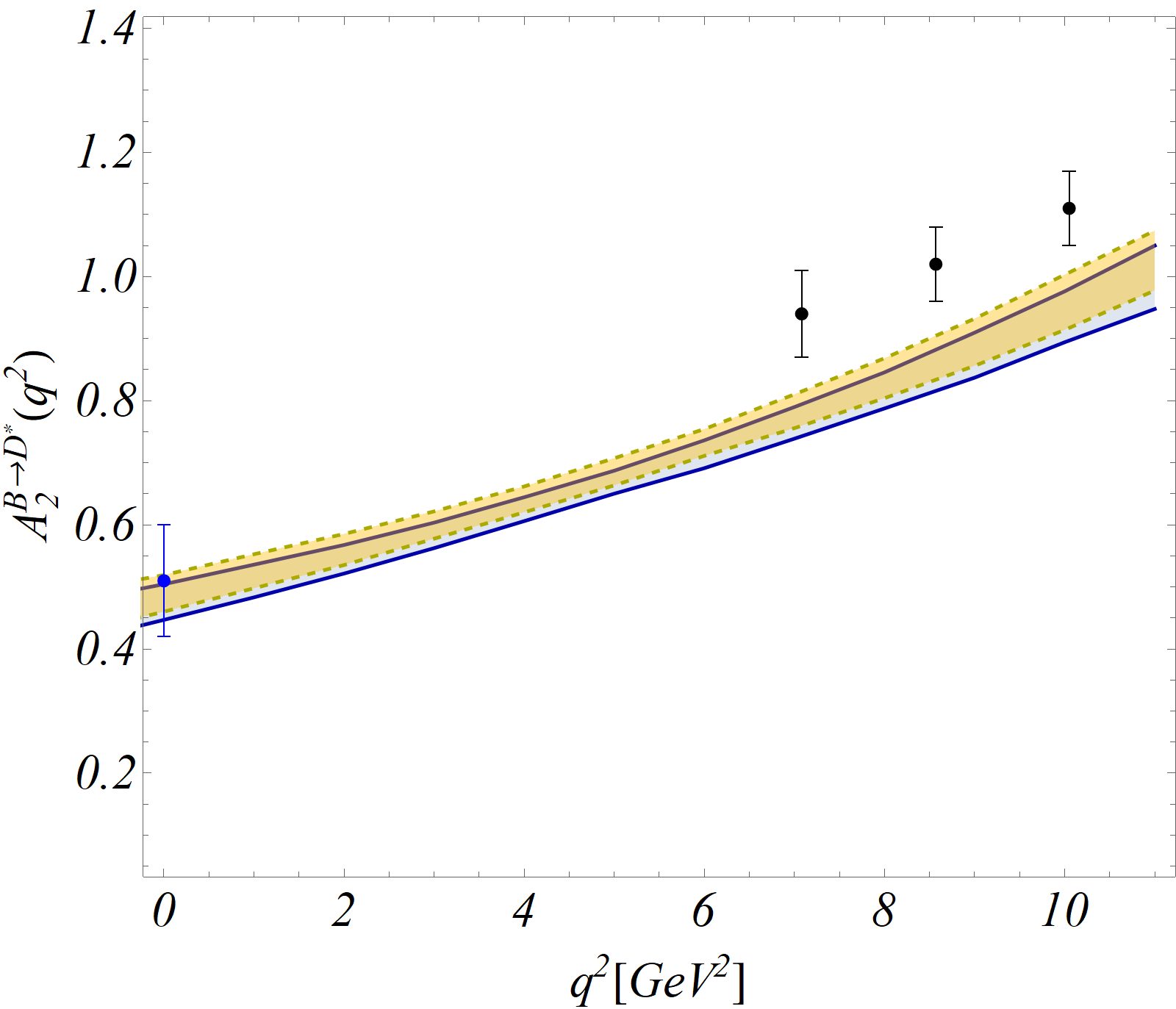}\label{fig:A2Dst}}~~~
	\subfloat[]{\includegraphics[width=0.3\textwidth]{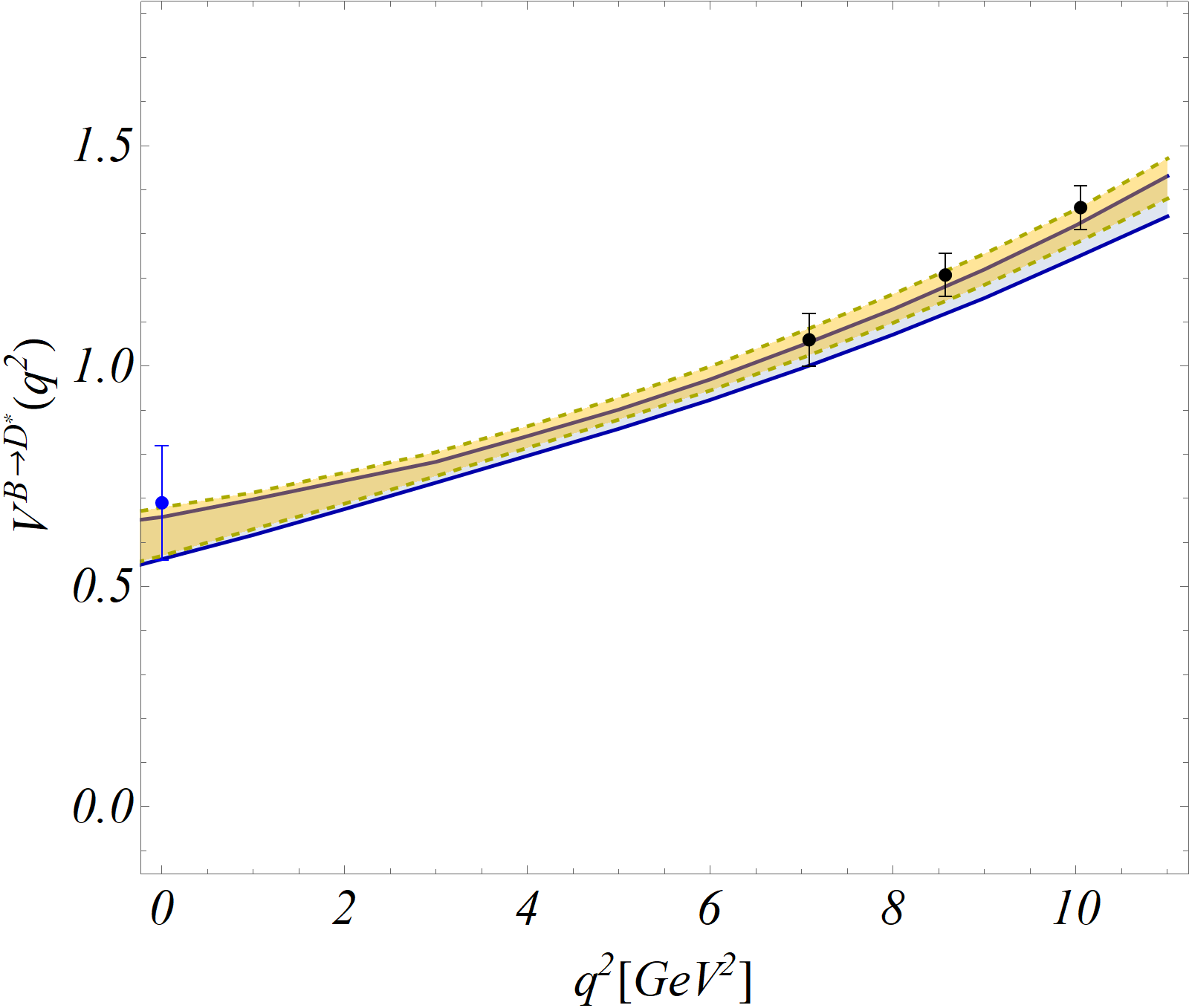}\label{fig:VDst}}\\
	\subfloat[]{\includegraphics[width=0.35\textwidth]{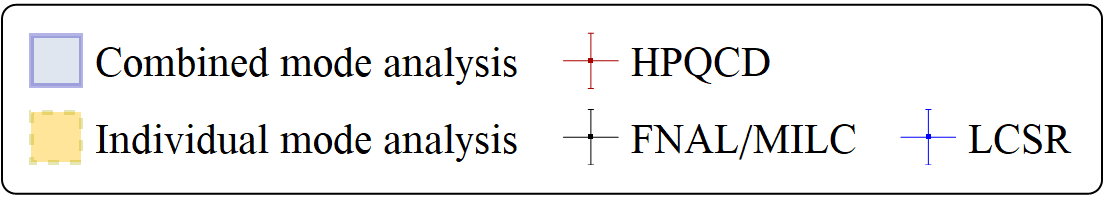}\label{fig:legb2c}}
	\caption{The $q^2$ distributions of the form factors in $\bar{B}\to D$ and $\bar{B}\to D^{*}$ decays. The lattice and LCSR datapoints are also shown in the plots. The LCSR inputs shown  are from \cite{Gubernari:2018wyi}.}
	\label{fig:FFB2D}
\end{figure*}
\begin{figure*}[t]
	\small
	\centering
	\subfloat[]{\includegraphics[width=0.35\textwidth]{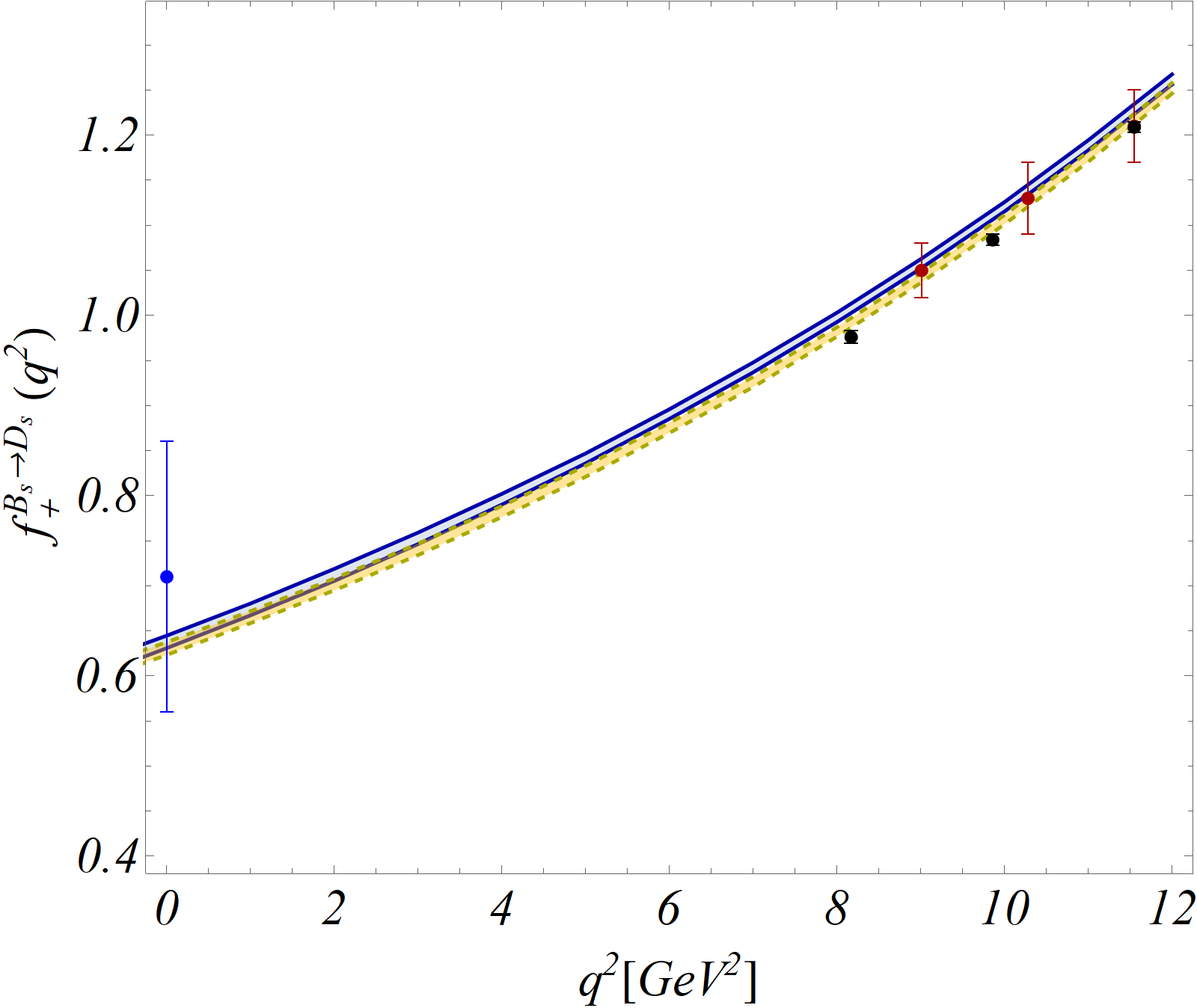}\label{fig:fplusBs2Ds}}~~~~~
	\subfloat[]{\includegraphics[width=0.35\textwidth]{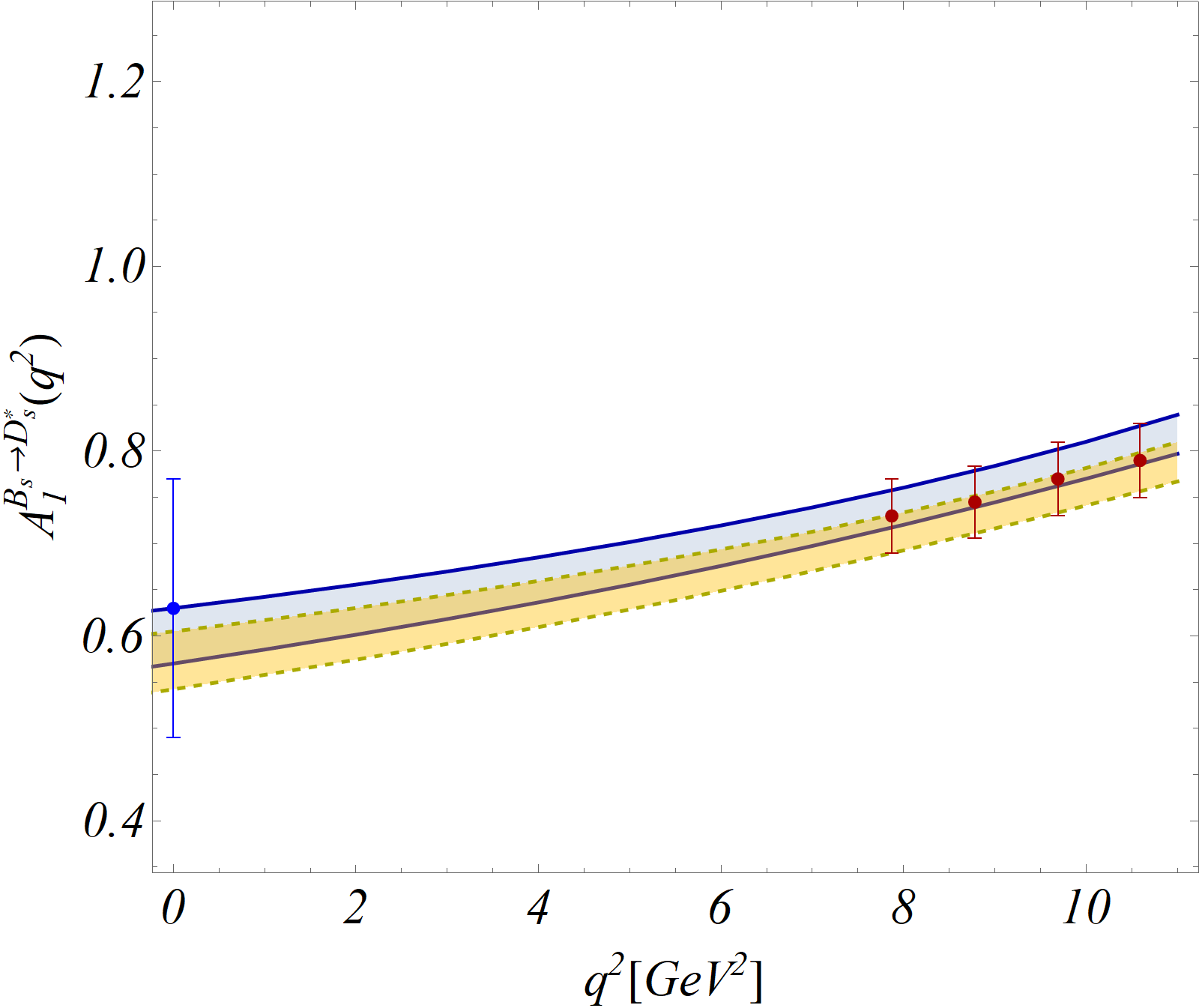}\label{fig:A1Dsst}}\\	\subfloat[]{\includegraphics[width=0.35\textwidth]{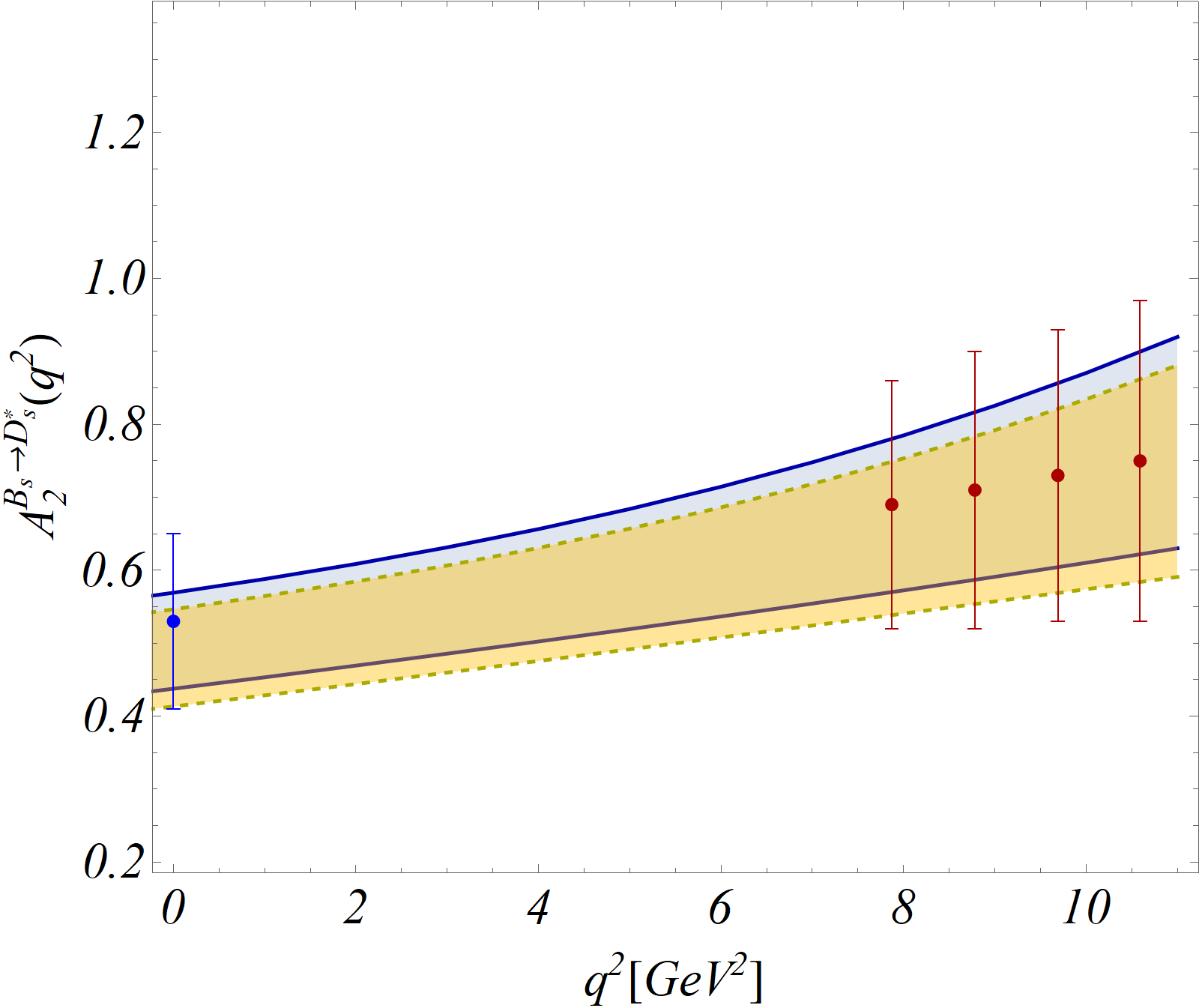}\label{fig:A2Dsst}}~~~~~
	\subfloat[]{\includegraphics[width=0.35\textwidth]{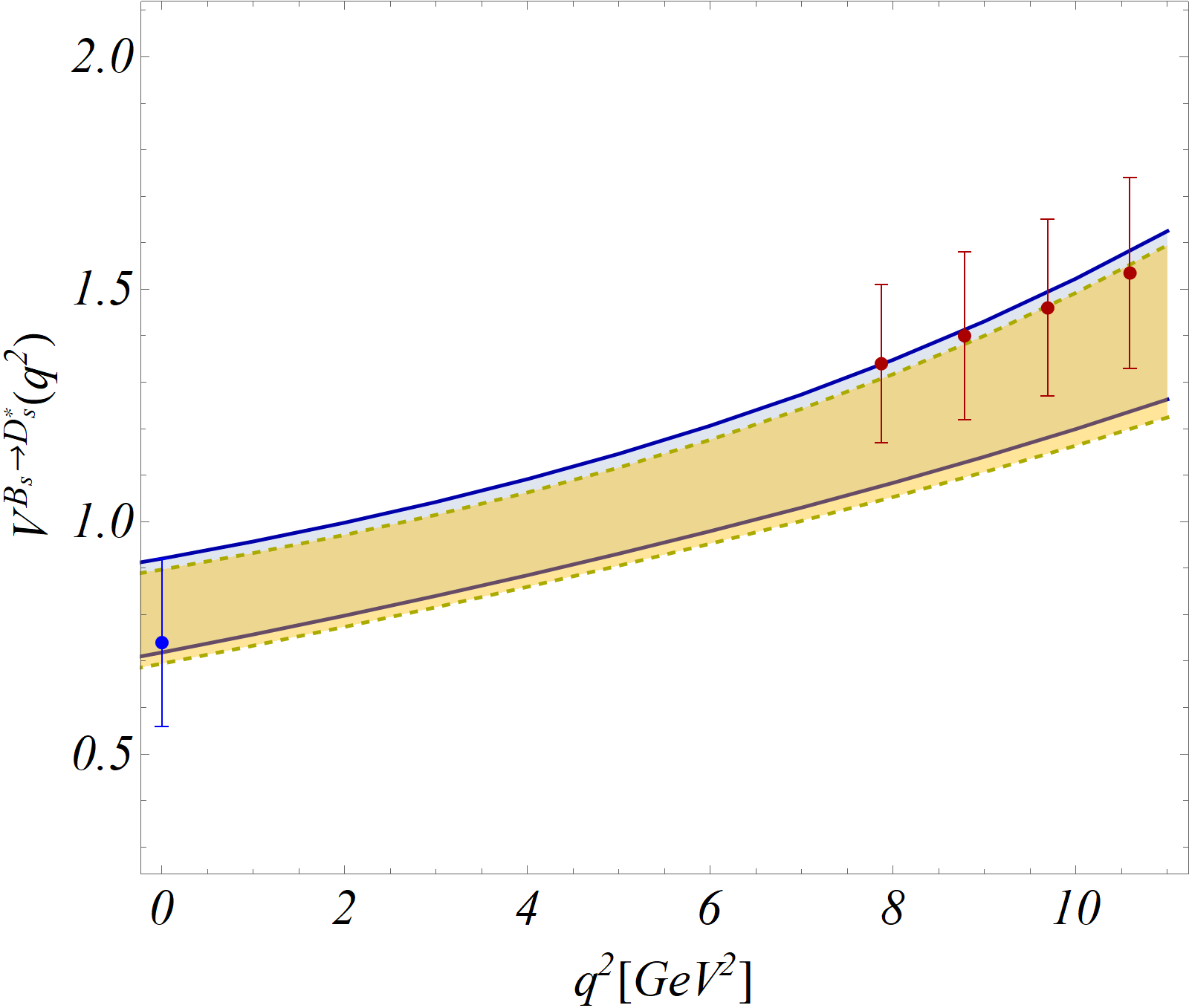}\label{fig:VDsst}}\\
	\caption{The $q^2$ distributions of the form factors in $\bar{B}_s\to D_s^{(*)}$ decays. The lattice and LCSR points are also shown in the plots. The legends are the same as that used in Fig.~\ref{fig:FFB2D}. The LCSR inputs shown are from \cite{Bordone:2019guc}.}
	\label{fig:FFBS2DS}
\end{figure*}

Apart from $\bar{B}_s \to D_s^{(*)}\ell^- \bar{\nu}_{\ell}$, experimental measurements are available on the partial branching fractions in bins of dilepton invariant mass squared $q^2$ for all the rest of the decays we mentioned above. In 2015, the Belle collaboration analyzed the $\bar{B}\to D l^- \bar{\nu}$ decay and provided measurements for the differential decay width in the recoil variable $w$ in 10 $w$-bins for both the charged and neutral $B$ decays with electrons and muons in the final state \cite{Belle:2015pkj}. In 2018 and 2023, the collaboration presented their results for the differential distributions in $w$, $\text{cos} \theta_l$, $\text{cos} \theta_v$ and $\chi$ in 10 bins for the $\bar{B}\to D^* l^- \bar{\nu}$ decay mode \cite{Belle:2018ezy,Belle:2023bwv}, which we have used in our analysis. In addition to these experimental results, we consider the inputs for the hadronic form factors available from various sources. For the $\bar{B} \to D$ form factors, we take lattice inputs from the Fermilab-MILC collaboration \cite{MILC:2015uhg} and the HPQCD collaboration \cite{Na:2015kha}. In \cite{MILC:2015uhg}, the form factors $f_+$ and $f_0$ are given at three values of the recoil variable $w$ = 1, 1.08 and 1.16. The HPQCD collaboration has provided the fit results for the form factor parameters following the BCL expansion, using which we have created synthetic data points for $f_+$ and $f_0$ at $w$ = 1, 1.06 and 1.12. The table for these inputs alongside the respective correlations can be obtained from refs.~\cite{Jaiswal:2017rve,Jaiswal:2020wer}. The lattice inputs for the form factors of $\bar{B} \to D^*$ mode are taken from the Fermilab-MILC collaboration \cite{FermilabLattice:2021cdg} where the values of the form factors $g, f, F_1$ and $F_2$ are provided for the first time at non-zero values of the recoil, namely $w$ = 1.03, 1.10 and 1.17 respectively. We use these pseudo data points and the given correlations directly in our analyses. The other relevant details of the method for analysing the data on $\bar{B}\to D^*\ell^- \bar{\nu}_{\ell}$ are similar to the one given in \cite{Jaiswal:2020wer}. Note that from the LCSR approach, for both the $\bar{B}\to D$ and $\bar{B}\to D^*$ form factors, we have incorporated the inputs only at $q^2=0$ given in ref.~\cite{Gubernari:2018wyi}.

In ref.~\cite{LHCb:2020cyw} the LHCb collaboration have extracted $|V_{cb}|$ by analysing the data and lattice inputs on $\bar{B}_s \to D_s^{(*)}\ell^- \bar{\nu}_{\ell}$ decays. Here, a different technique is introduced for the measurement of $|V_{cb}|$. Instead of probing the rate distributions over the $q^2$ or the recoil angles, another variable fully reconstructed from the visible decay products is used. This variable is the momentum of the $D_{s}^{(*)}$ meson transverse to the $B_s$ flight direction, called $P_{\perp}(D_s)$. Note that for $\bar{B}_s\to D_s^{*+}\mu^- \bar{\nu}$ decays, $D_s^*$ is reconstructed via its decay $D_s^*\to D_s X$ (with $X = \gamma$ or $\pi^0$), and hence the momentum of $D_s^*$ is not known. For this case too, the $P_{\perp}(D_s)$ is considered as the component of the $D_s$ momentum transverse to  flight direction of the $B_s$ meson. Neither does the paper provide any relation between the $P_{\perp}(D_s)$ and the recoil angle, nor can their results be directly used in other phenomenological analyses. However, they do provide the fit results for the BGL coefficients along with some other inputs which can be useful to create pseudo data for the rate with the recoil angle ($w$). In the LHCb paper \cite{LHCb:2020cyw}, the signal yield $N_{sig}$ is defined as :
\begin{equation}
N_{sig}^{(*)} = \mathcal{N}^{(*)} \tau \int \frac{d \Gamma (\bar{B}_s \to D_s^{(*)} \mu^- \bar{\nu})} {d \zeta} d \zeta
\end{equation}
where $\zeta$ could be the recoil angle $w$ for $\bar{B}_s \to D_s \mu^- \bar{\nu}$ and $(w, cos \theta_{\mu}, cos \theta_D, \chi)$ for $\bar{B}_s \to D_s^* \mu^- \bar{\nu}$ channels. Here, the other parameters are defined as: 
\begin{eqnarray}\label{eq:lhcbotherinputs}
\mathcal{N}^{(*)} &=& \frac{N_{ref}^{(*)} \xi^{(*)} K^{(*)}} {B(B^0 \to D^{(*)-} \mu^+ \nu_{\mu})} \, \\
K &=& \frac{f_s}{f_d} \frac{B(D_s^- \to K^+ K^- \pi^-)} {B(D^- \to K^+ K^- \pi^-)} \, \\
K^* &=& \frac{f_s}{f_d} \frac{B(D_s^- \to K^+ K^- \pi^-)} {B(D^{*-} \to D^- X) B(D^- \to K^+ K^- \pi^-)}
\end{eqnarray}
where $f_s/f_d$ is the ratio of $B_s^0$ to $B^0$ meson production fractions. Details of these inputs are provided in the LHCb paper. Using the fit results of table-IX of the LHCb paper \cite{LHCb:2020cyw} and the respective correlations, we have created the pseudo data for the signal events in small $w$-bins, which we have presented in table \ref{tab:syntheticlhcb}. Note that for $B_s \to D_s$ transitions, there are existing results from Lattice groups at non-zero recoil, namely Fermilab-MILC \cite{Bailey:2012rr} and HPQCD \cite{McLean:2019qcx} where the fit results of the respective form factor parameters are provided following BGL and BCL parametrizations respectively. For $B_s \to D_s^*$ decays, the HPQCD collaboration has given pseudo data for the form factors at $w = 1, 1.04, 1.08$, and $1.12$ \cite{Harrison:2021tol}. For the $B_s\to D_s$ form factors, LHCb has used the fit results for the BGL coefficients from ref.~\cite{McLean:2019qcx} as nuisance parameters in their fits, while for the $B_s \to D_s^*$ form factors, they have used only the input from HPQCD on $f^s(w=1)$ \cite{McLean:2019sds} which was available by then. They have not considered any input from the Fermilab-MILC collaboration.

As a cross check, we have reproduced the fit results (with the identical inputs) as given in the LHCb paper using the pseudo data points from table \ref{tab:syntheticlhcb}. For comparison, we have presented the result of this fit in table \ref{tab:lhcbfitresult} alongside the fit results of LHCb. The results are absolutely consistent with each other. We have fixed the values of a few other inputs defined in eq.~\ref{eq:lhcbotherinputs} to their central values in this fit, since the associated errors have negligible impact on the extraction of $|V_{cb}|$. Note that  LHCb carried out their analysis with the mass of the muon $m_{\mu} = 0$. In this analysis for $\bar{B}_s \to D_s^{(*)}\mu^- \bar{\nu}_{\mu}$ decays we have set the mass of the muon to zero. The extracted value of $|V_{cb}|$ is not sensitive to the muon mass which has been observed earlier in the studies of $\bar{B}\to D^{(*)}\mu^- \bar{\nu}_{\mu}$ decays \cite{Abdesselam:2017kjf,Belle:2018ezy}.

We use the pseudo data points given in table \ref{tab:syntheticlhcb} as inputs alongside the lattice inputs on the form factors for both the decays in the zero ($w=1$) and non-zero ($w\ne 1$) recoil regions from the refs.~\cite{Bailey:2012rr,McLean:2019qcx,Harrison:2021tol}. For $B_s \to D_s^*$ decays, we have directly used the pseudo data for the form factors at $w = 1, 1.04, 1.08$, and $1.12$ provided by the HPQCD collaboration \cite{Harrison:2021tol}. For $B_s\to D_s$ decays, we have generated correlated synthetic data points for the form factors $f_+$ and $f_0$ using the fit results of the Fermilab-MILC analysis \cite{Bailey:2012rr} at three values of the recoil variable, namely $w$ = 1, 1.08 and 1.16 and from the fit results of the HPQCD analysis \cite{McLean:2019qcx} at $w$ = 1, 1.06 and 1.12\footnote{There is no particular reason behind choosing the specified values of $w$ to create the pseudo data. For $B_s\to D_s$ form factors, Fermilab/MILC has provided the results for the BGL coefficients after fitting the pseudo data created at the points $w$ = 1, 1.08 and 1.16 \cite{Bailey:2012rr}. We simply chose the same points to create the pseudo data. For the pseudo data corresponding to the HPQCD fit, we did not choose points beyond $w=1.12$, as their results are directly calculated for $q^2 \gsim 9.5$ GeV$^2$ or $w \lsim 1.11$ \cite{Harrison:2021tol}. We note that the lattice data points they have provided for the $B_s \to D_s^*$ form factors contain $w = 1, 1.04, 1.08$, and $1.12$, and none beyond. A similar argument applies to $B\to D$ form factors for HPQCD. }. In addition, we have incorporated the LCSR inputs for all the form factors only at $q^2=0$ from ref.~\cite{Bordone:2019guc}. We truncate the series expansion given in eq.~\ref{eq:FF-BGL} at $N=2$ and note that the extracted value of $|V_{cb}|$ does not change for the fit with $N=3$. However, for that fit, the higher-order coefficients $a_3^g$ of $g(q^2)$ has a large error and goes beyond the unitarity bound. Hence this coefficient is mostly unconstrained, which means one has to wait for more precise data.   

In table \ref{tab:Vcb}, we have presented our fit results for $|V_{cb}|$ in all the different analyses. The fit results for the coefficients along with the respective correlations are provided separately in a json file. In all the fit results we have discussed in this paper, the respective errors are presented at their 1-$\sigma$ confidence interval (CL). As described above, the pseudo data obtained on the decay rates for $B_s\to D_s$ and $B_s\to D_s^*$ are correlated. Hence, we have not done separate analyses for these two modes. The shape of the form factors as a function of $q^2$ in both $B$ and $B_s$ decays to $D^{(*)}$ and $D_s^{(*)}$ are shown in Figs.~\ref{fig:FFB2D} and \ref{fig:FFBS2DS}, respectively. To see the impact of the combined analysis, we have plotted the shapes of the form factors from the combined analysis against those from the analyses of the individual modes. We see that they are pretty consistent with each other. We have not shown the shape for $f_0^{B_s\to D_s}$ and $A_0^{B_s\to D_s^*}$ since we have not constrained them from the fits. This is because we have obtained the decay rates for $B_s$ decays with $m_{\mu} = 0$. Hence the respective rates are not sensitive to these form factors. 

The extracted value of $|V_{cb}|$ is lower as obtained from $B$ decays than from $B_s$ decays. In general, the inputs from Fermilab/MILC are very precise. They play an essential role in controlling the overall error of the fit parameters, which are also visible in the obtained shape of the form factors wherever applicable; for example, see figure \ref{fig:FFB2D} and \ref{fig:fplusBs2Ds}. Also, we observe that the inclusion of the inputs on $f_+^{B_s\to D_s}$ from the Fermilab/MILC results in a slight upward shift in the value of $|V_{cb}|$ as compared to the fit without these inputs. The extracted value of $|V_{cb}|$ from a combined fit, including all the available inputs, is given by 
\begin{equation}\label{eq:Vcbpred}
|V_{cb}| = (40.5 \pm 0.6)\times 10^{-3}
\end{equation}
whose deviations from the inclusive determination of (41.69 $\pm$ 0.63) $\times$ $10^{-3}$ obtained from the analyses of the $q^2$ moments and the differential rates \cite{Bernlochner:2022ucr}, and from the inclusive determination of (42.16 $\pm$ 0.51) $\times$ $10^{-3}$ obtained using lepton energy and hadronic invariant mass moments distributions \cite{Bordone:2021oof} are $\sim$ 1.4 and 2.1 $\sigma$ respectively. Note that we have obtained a very low p-value of this combined fit. We have checked that the reason for such a low fit probability is the lattice datapoints on $f_{+,0}^{B_s\to D_s}$ from the Fermilab/MILC collaboration \cite{Bailey:2012rr}\footnote{In particular, the points at $w$ = 1.08 and 1.16 are picked up as outliers with pull $\sim$ 4.}. On dropping these data points, the p-value of the fit increases significantly ($\sim 21.8 \%$ ) and $|V_{cb}|$ = (40.3 $\pm$ 0.6) $\times$ $10^{-3}$, which is consistent with the result of the combined fit shown in eq.~\ref{eq:Vcbpred}. The corresponding fit result is in the bottom row of table \ref{tab:Vcb}.The dropping of 
	the Fermilab/MILC data for $B_s\to D_s$ transitions is motivated by the
	lack of a complete analysis of the systematic uncertainties in these form
	factors. 

Following the dispersive matrix approach in the description of the hadronic form factors, the extracted value $|V_{cb}| = (41.2 \pm 0.8)\times 10^{-3}$ which has been obtained from an average of the extracted values of $|V_{cb}|$ from the $\bar{B}\to D^{(*)}\ell^- \bar{\nu}_{\ell}$ and $\bar{B}_s\to D_s^{(*)}\ell^- \bar{\nu}_{\ell}$ \cite{Martinelli:2021myh,Martinelli:2021onb,Martinelli:2022xir}. This is very much consistent with the results we have obtained from a combined fit of all the available inputs. 

Apart from $A_2^{B\to D^*}(q^2)$, the $q^2$ shapes of almost all the other form factors obtained from the fit results are more or less consistent with the lattice or LCSR inputs. From Fig.~\ref{fig:A2Dst} we can see that the shapes of $A_2^{B\to D^*}(q^2)$ obtained from fits (individual and combined) are not consistent (at 1-$\sigma$) with the lattice inputs from Fermilab/MILC at $w$ = 1.03, 1.10 and 1.17 respectively, while the $q^2$ shapes of the rest of the form factors like $f_{+,0}^{B\to D}(q^2)$, $A_{0,1}^{B\to D^*}(q^2)$ and $V^{B\to D^*}(q^2)$ are consistent with the lattice inputs from Fermilab/MILC and wherever applicable with HPQCD (Fig.~\ref{fig:FFB2D}). Similarly, it is evident from Fig.~\ref{fig:FFBS2DS} that the shape of all the form factors related to $B_s\to D_s^{(*)}$ decays are fully consistent with the HPQCD estimates. However, we note from Fig.~\ref{fig:fplusBs2Ds} that the pseudo data points for $f_+^{B_s\to D_s}$ from Fermilab/MILC are very precise and lie on the edge of the $q^2$ distribution determined from the fits. In particular, the point at $w=1.16$ lies outside the fitted region. The LCSR predictions at $q^2=0$ have large errors and are entirely consistent with the fit results with negligible impact on the fits.                    

\begin{table}[t]
	\begin{center}
		\scriptsize
		\renewcommand*{\arraystretch}{1.6}
		\begin{tabular}{|c|c|c|c|c|c|}
			\hline
			Mode  & Fits with the Inputs & $\chi^2_{min}$/DOF& p-value(\%) & $|V_{ub}|$ $\times$ $10^3$ & $|V_{ub}|/|V_{cb}|$\\
			\hline
			$\bar{B} \to \pi l^- \bar{\nu}$  &(I)  All Experiment+JLQCD  & 10.8/11 & 45.7  & 3.98 $\pm$ 0.42 & 0.098 $\pm$ 0.010 \\
		  & (II) Fit-I + LCSR \cite{Leljak:2021vte,Gubernari:2018wyi}  & 16.3/29 & 97.2 & 3.73 $\pm$ 0.24 & 0.092 $\pm$ 0.006\\
		 & (III)	All Experiment+Fermilab/MILC  & 11.7/12 & 46.6 & 3.78 $\pm$ 0.16 & 0.093 $\pm$ 0.004\\
		& (IV)	Fit-III + LCSR \cite{Leljak:2021vte,Gubernari:2018wyi}  & 24.2/30 & 76.2 & 3.84 $\pm$ 0.15 & 0.095 $\pm$ 0.004\\
		& (V)	All Experiment+RBC/UKQCD & 11.0/11  & 44.3 & 3.75 $\pm$ 0.34 & 0.093 $\pm$ 0.009\\
		&(VI) Fit-V + LCSR \cite{Leljak:2021vte,Gubernari:2018wyi} & 17.5/29 & 95.3 & 3.57 $\pm$ 0.23 & 0.088 $\pm$ 0.006\\
			\cline{2-6}
		&(VII)	  All Experiment+ & 27.2/17 & 5.5 & 3.69 $\pm$ 0.24 & 0.091 $\pm$ 0.006	\\
		& RBC/UKQCD+JLQCD & & & & \\
		&(VIII)	  All Experiment+ All Lattice & 50.8/24 & 0.1 & 3.69 $\pm$ 0.13 & 0.091 $\pm$ 0.003\\
		& (IX)	Fit-VIII + LCSR \cite{Leljak:2021vte,Gubernari:2018wyi}  & 55.7/42 & 7.7 & 3.69 $\pm$ 0.12 & 0.091 $\pm$ 0.003 \\
			\hline 
			$\bar{B} \to \pi l^- \bar{\nu}$ and & (X)  All Experiment + All Lattice  & 74.9/38  & 0.03 & 3.60 $\pm$ 0.11 & 0.089 $\pm$ 0.003\\
			$\bar{B}_s \to K l^- \bar{\nu}$ 
			& (XI) Fit-(X) + LCSR \cite{Gubernari:2018wyi,Leljak:2021vte,Khodjamirian:2017fxg} & 95.5/57 & 0.1 & 3.50 $\pm$ 0.10 & 0.086 $\pm$ 0.003\\
			& (XII) All Experiment+ All Lattice & 58.8/32 & 0.3  & 3.74 $\pm$ 0.12 & 0.092 $\pm$ 0.003\\
			 & except inputs from  & &  & & \\  
			 & HPQCD on $B_s \to K$ \cite{HPQCD:BStoK2014} & & & & \\
		 &  (XIII) Fit-(XII) + LCSR \cite{Gubernari:2018wyi,Leljak:2021vte,Khodjamirian:2017fxg} & 88.4/51 & 0.1 & 3.52 $\pm$ 0.10 & 0.087 $\pm$ 0.003\\
			\hline
			$\bar{B} \to \rho l^- \bar{\nu}$ &  Experiment + LCSR \cite{PhysRevD.101.074035,Gubernari:2018wyi,Straub:2015ica} & 31.2/41 & 86.6 & 3.22 $\pm$ 0.26 & 0.080 $\pm$ 0.007\\
			\hline
			$\bar{B} \to \omega l^- \bar{\nu}$ &  Experiment+LCSR \cite{PhysRevD.101.074035,Straub:2015ica} & 8.6/15 & 89.7 & 3.09 $\pm$ 0.33 & 0.076 $\pm$ 0.008\\
			\hline
			$\bar{B} \to \pi(\rho,\omega) l^- \bar{\nu}$ &  All inputs combined & 99.6/100 & 49.4 & 3.58 $\pm$ 0.11 & 0.088 $\pm$ 0.003 \\
			\hline
			&  $B\to \pi$ and $B_s \to K$:  & 117.8/96 & 6.5 & $3.52 \pm 0.10$ & $0.087 \pm 0.003$\\
			& Experiment + Lattice & & & &\\
	$\bar{B} \to \pi(\rho,\omega) l^- \bar{\nu}$ and		&   $B \to (\rho,\omega)$: Experiment   & & & &\\
	&  + LCSR & & & &\\
		\cline{2-6}
		$\bar{B}_s \to K l^- \bar{\nu}$	  & For all modes: Experiment + & 137.2/115 & 7.7 & 3.45 $\pm$ 0.09 & 0.085 $\pm$ 0.003 \\
			 &  lattice + LCSR & & & &\\
			\hline
		\end{tabular}
		\caption{A comparative study of the impact of the form factor inputs for different channels on the extraction of $|V_{ub}|$. }
		\label{tab:Vub}
	\end{center}
\end{table}

  	\begin{figure*}[t]
  		\small
  		\centering
  		\subfloat[]{\includegraphics[width=0.35\textwidth]{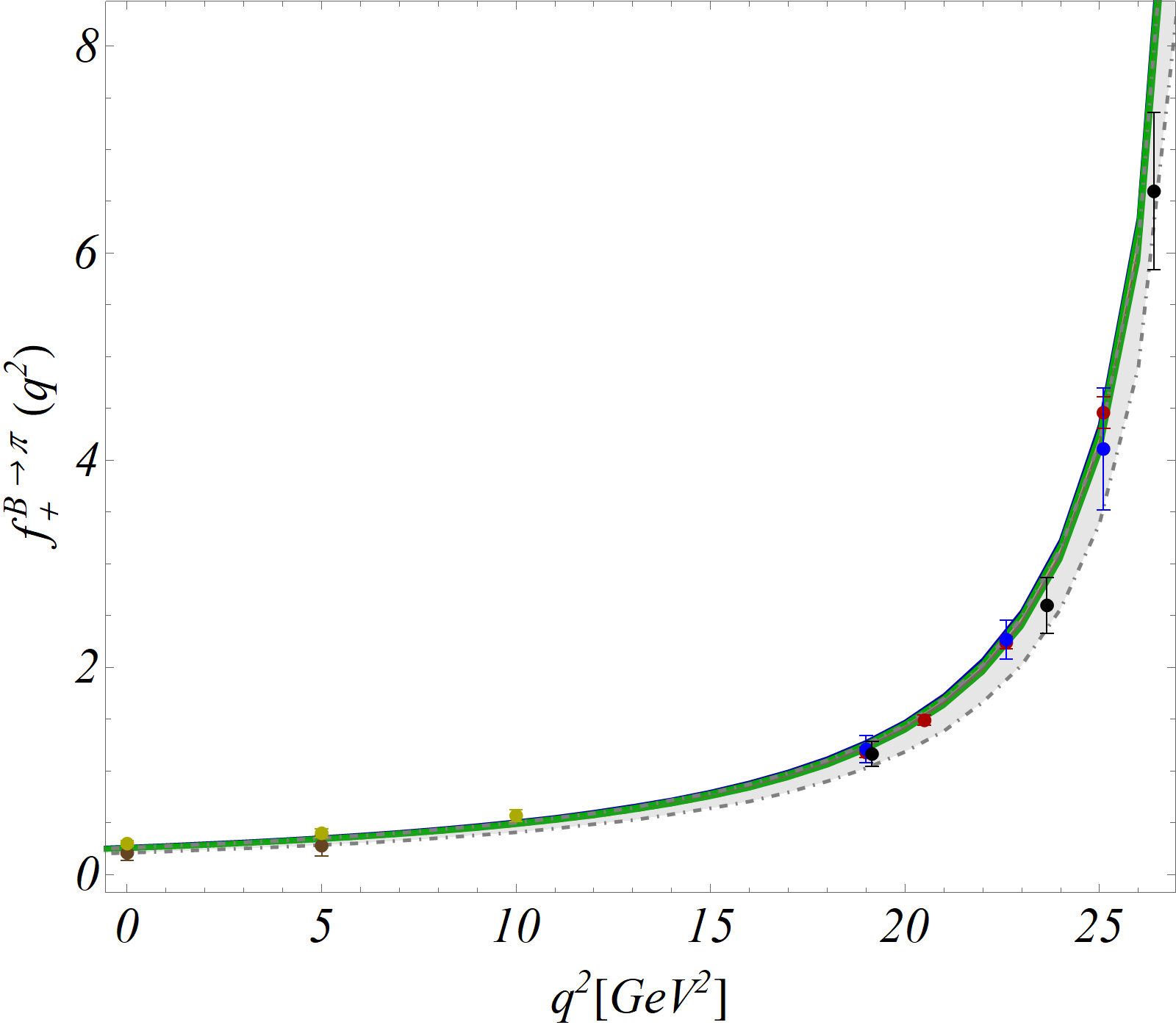}\label{fig:fplus}}~~~~~
  		\subfloat[]{\includegraphics[width=0.36\textwidth]{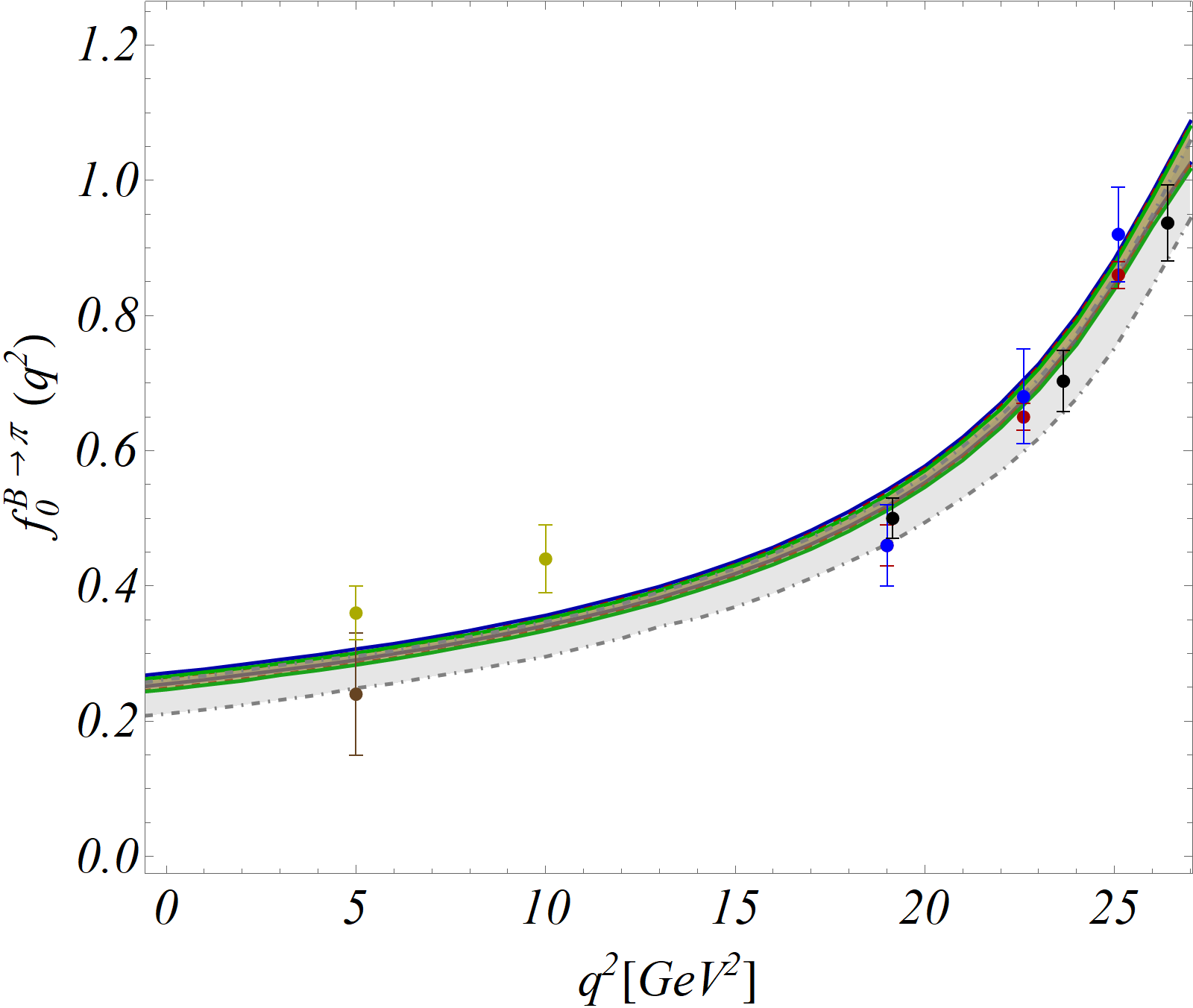}\label{fig:f0}}\\
  		\subfloat[]{\includegraphics[width=0.45\textwidth]{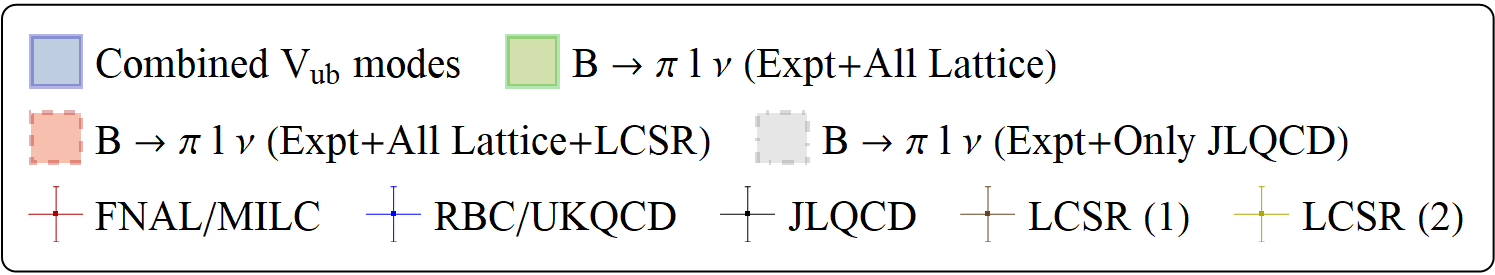}\label{fig:legpi}}
  		\caption{The $q^2$ distributions of the form factors $f_+$ and $f_0$ for $\bar{B} \to \pi l^- \bar{\nu}$ modes. The lattice and LCSR datapoints are also shown in the plots. The labels ``LCSR(1)" and ``LCSR(2)" refer to the inputs from LCSR \cite{Gubernari:2018wyi} and \cite{Leljak:2021vte} respectively.}
  		\label{fig:FFpi}
  	\end{figure*}
  	
  	\begin{figure*}[t]
  		\small
  		\centering
  		\subfloat[]{\includegraphics[width=0.35\textwidth]{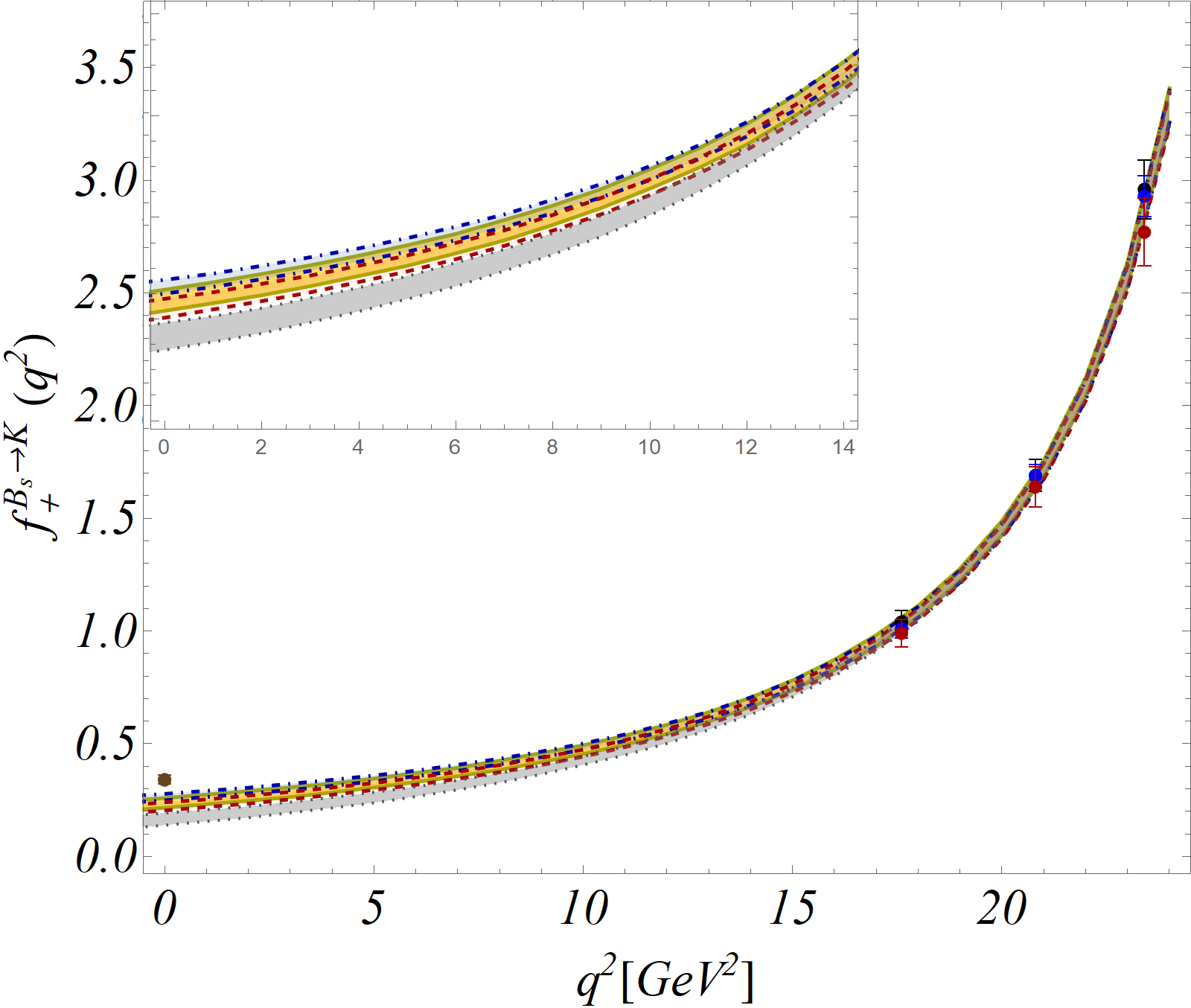}\label{fig:fplusBsK}}~~~~~
  		\subfloat[]{\includegraphics[width=0.36\textwidth]{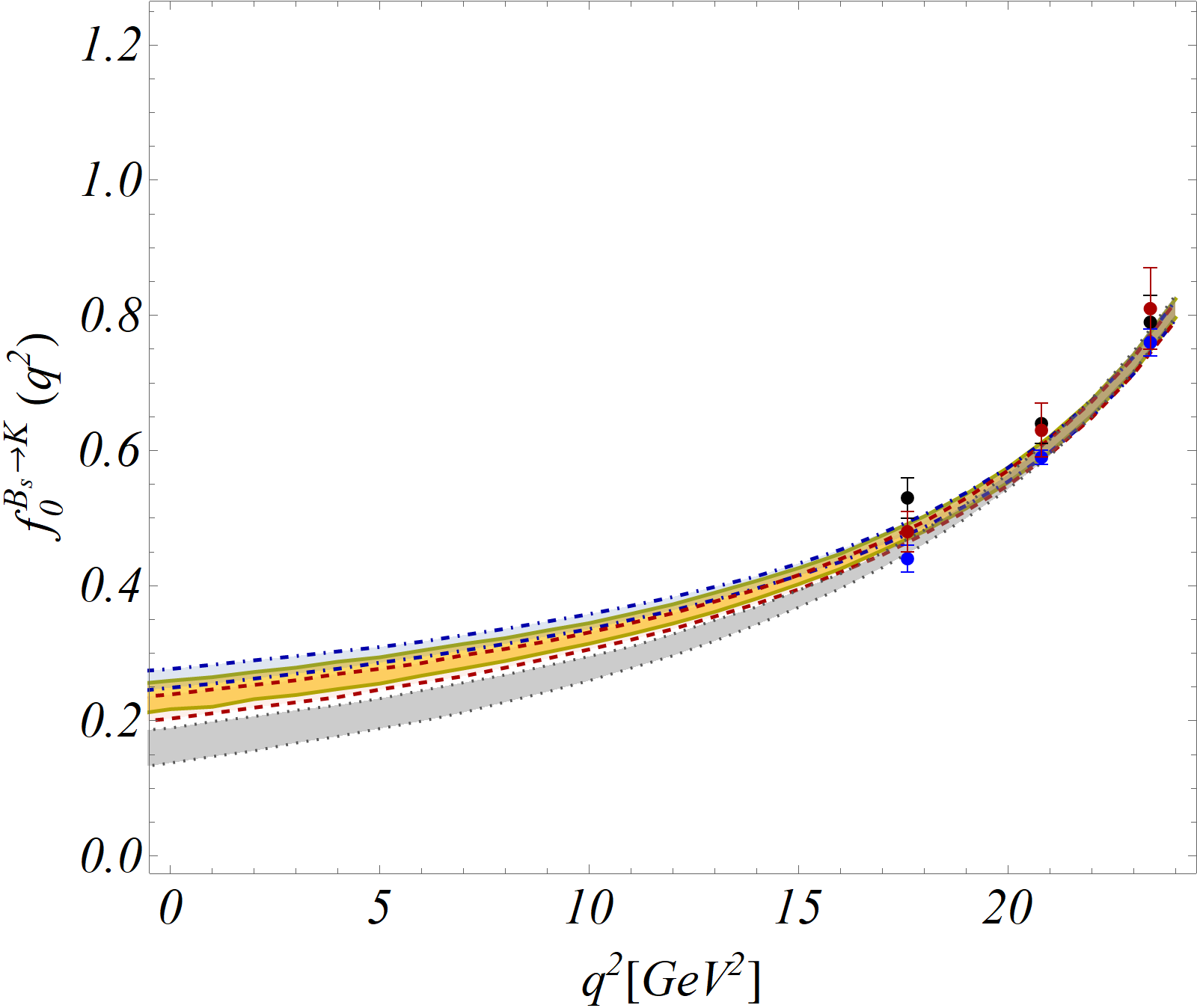}\label{fig:f0BsK}}\\
  		\subfloat[]{\includegraphics[width=0.65\textwidth]{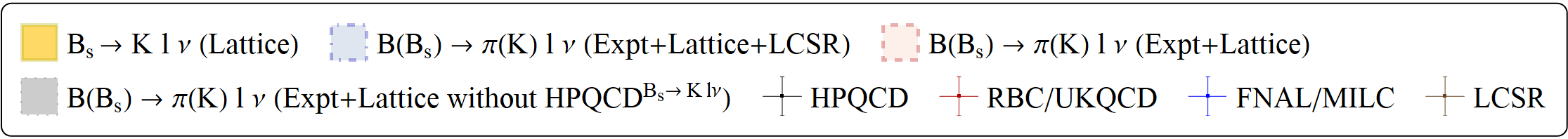}\label{fig:legBsK}}
  		\caption{The $q^2$ distributions of the form factors $f_+$ and $f_0$ for $\bar{B}_s \to K \mu^- \bar{\nu}_{\mu}$ modes. The lattice and LCSR datapoints are also shown in the plots. }
  		\label{fig:FFBsK}
  	\end{figure*}
  		\begin{figure*}[t]
  			\small
  			\centering
  			\subfloat[]{\includegraphics[width=0.3\textwidth]{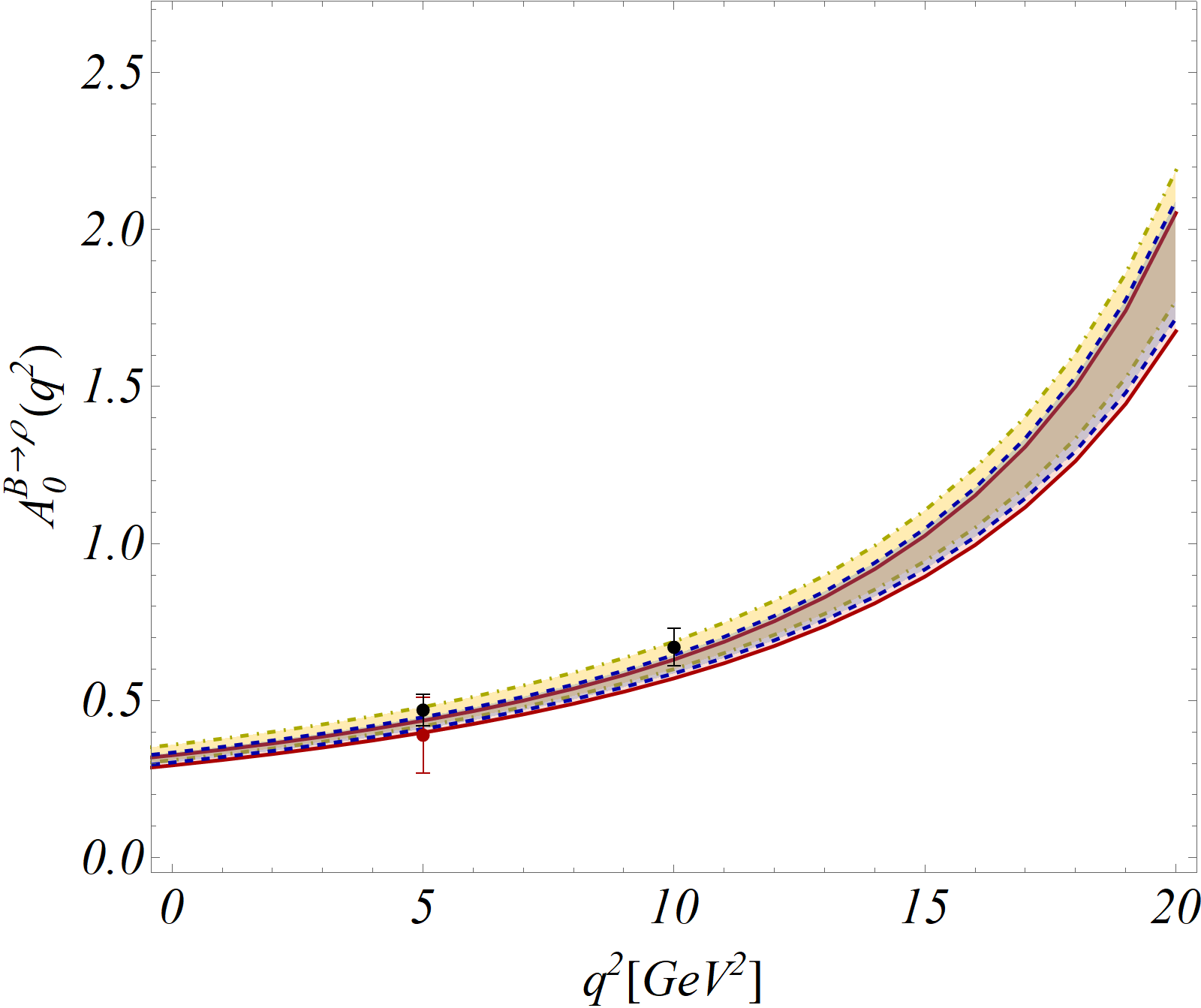}\label{fig:A0Rho}} ~~~~
  			\subfloat[]{\includegraphics[width=0.3\textwidth]{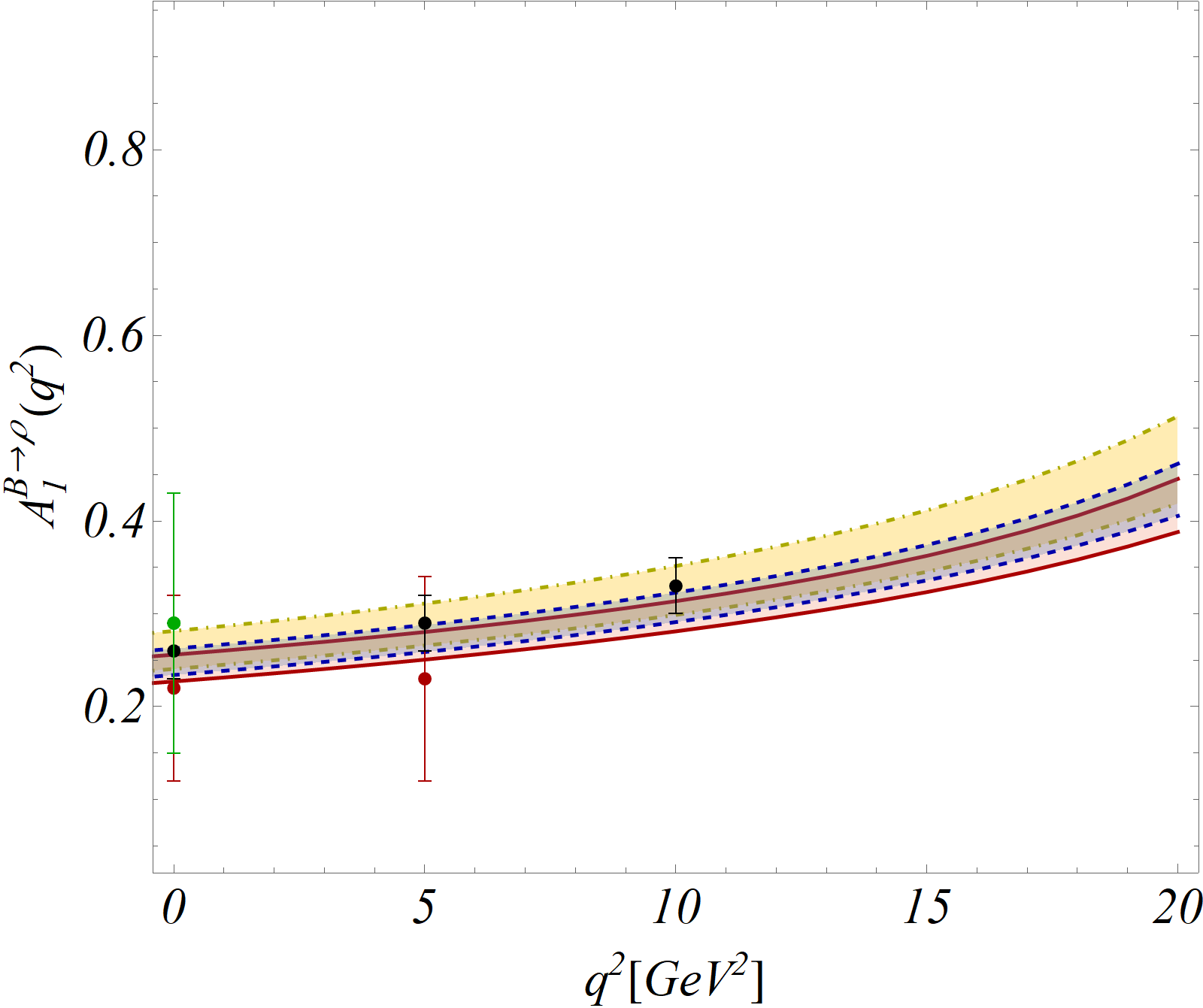}\label{fig:A1Rho}}~~~~	\subfloat[]{\includegraphics[width=0.3\textwidth]{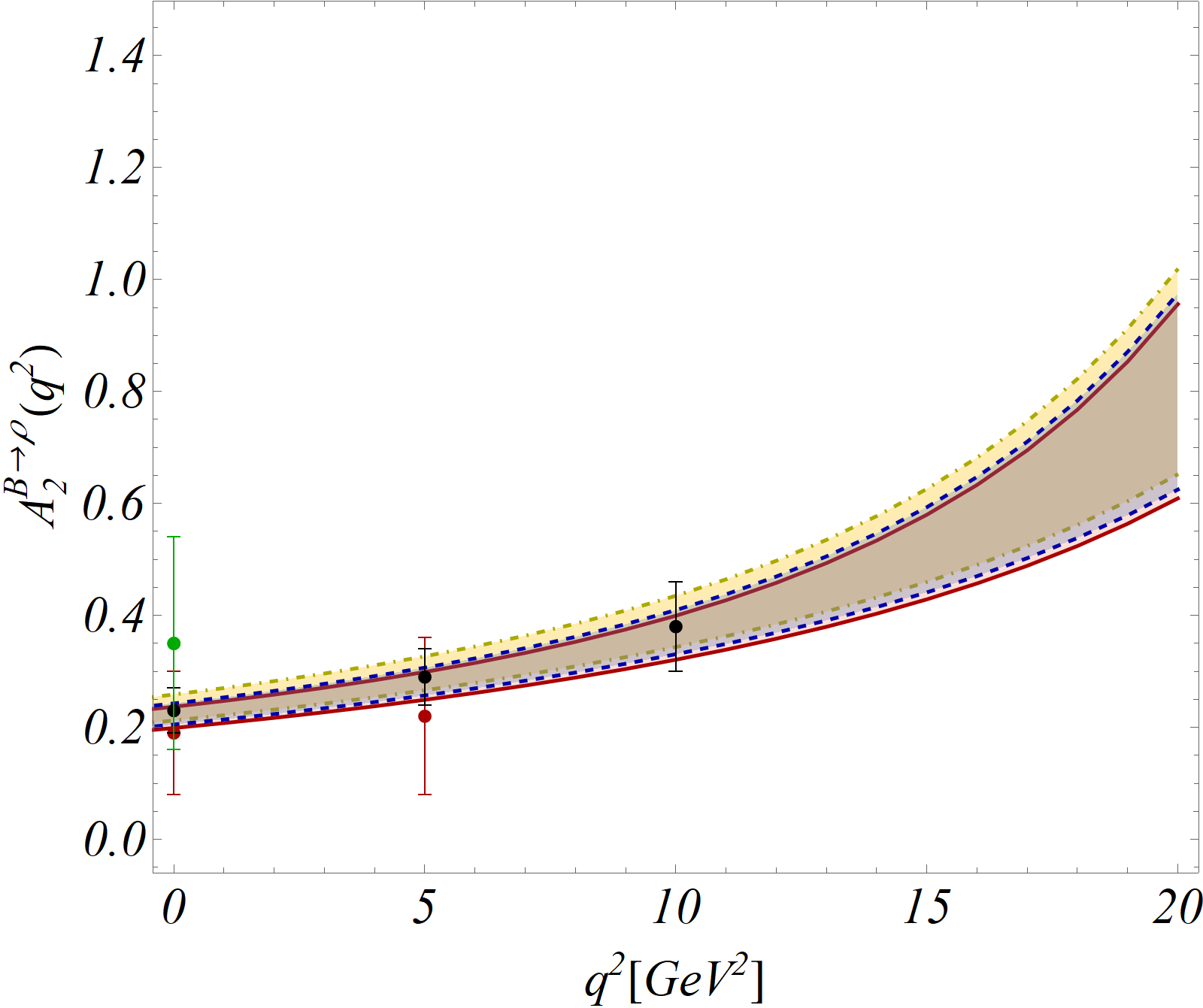}\label{fig:A2Rho}} \\
  			\subfloat[]{\includegraphics[width=0.3\textwidth]{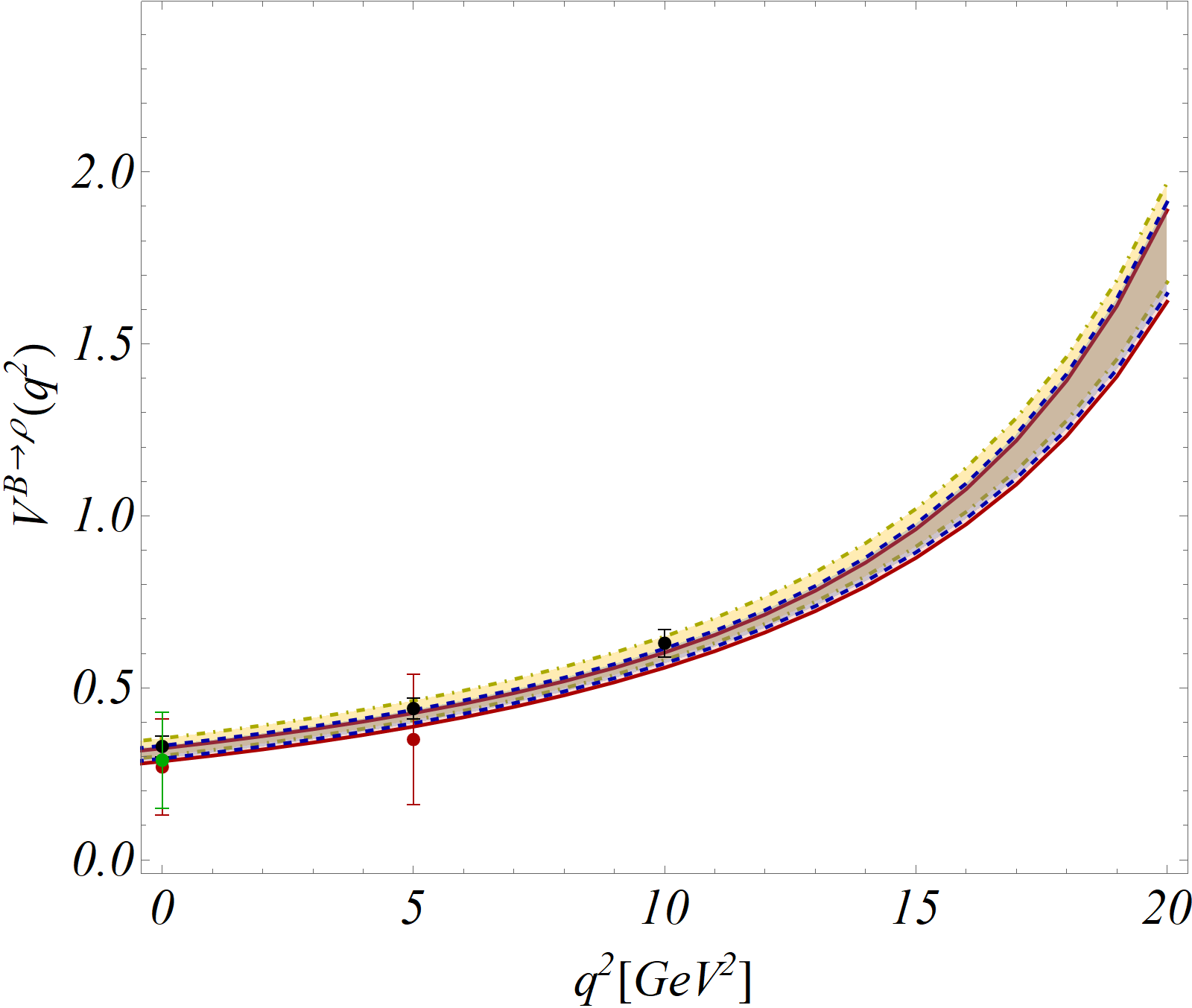}\label{fig:VRho}}~~~~
  			\subfloat[]{\includegraphics[width=0.25\textwidth]{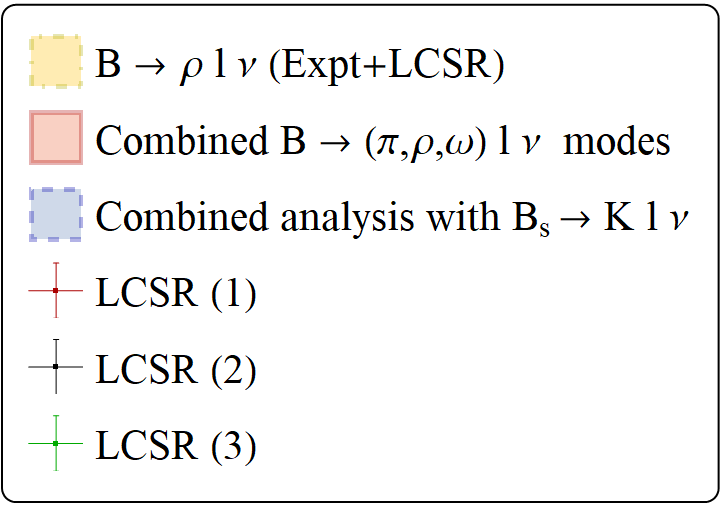}\label{fig:legRho}}
  			\caption{The $q^2$ distributions of the form factors $A_0$, $A_1$, $A_2$ and $V$ for $\bar{B} \to \rho l^- \bar{\nu}$ modes. The LCSR datapoints are also shown in the plots. The labels ``LCSR(1)", ``LCSR(2)" and ``LCSR(3)" refer to the inputs from LCSR \cite{Gubernari:2018wyi}, \cite{Straub:2015ica} and \cite{PhysRevD.101.074035} respectively.}
  			\label{fig:FFrho}
  		\end{figure*}
  		
  		\begin{figure*}[t]
  			\small
  			\centering
  			\subfloat[]{\includegraphics[width=0.3\textwidth]{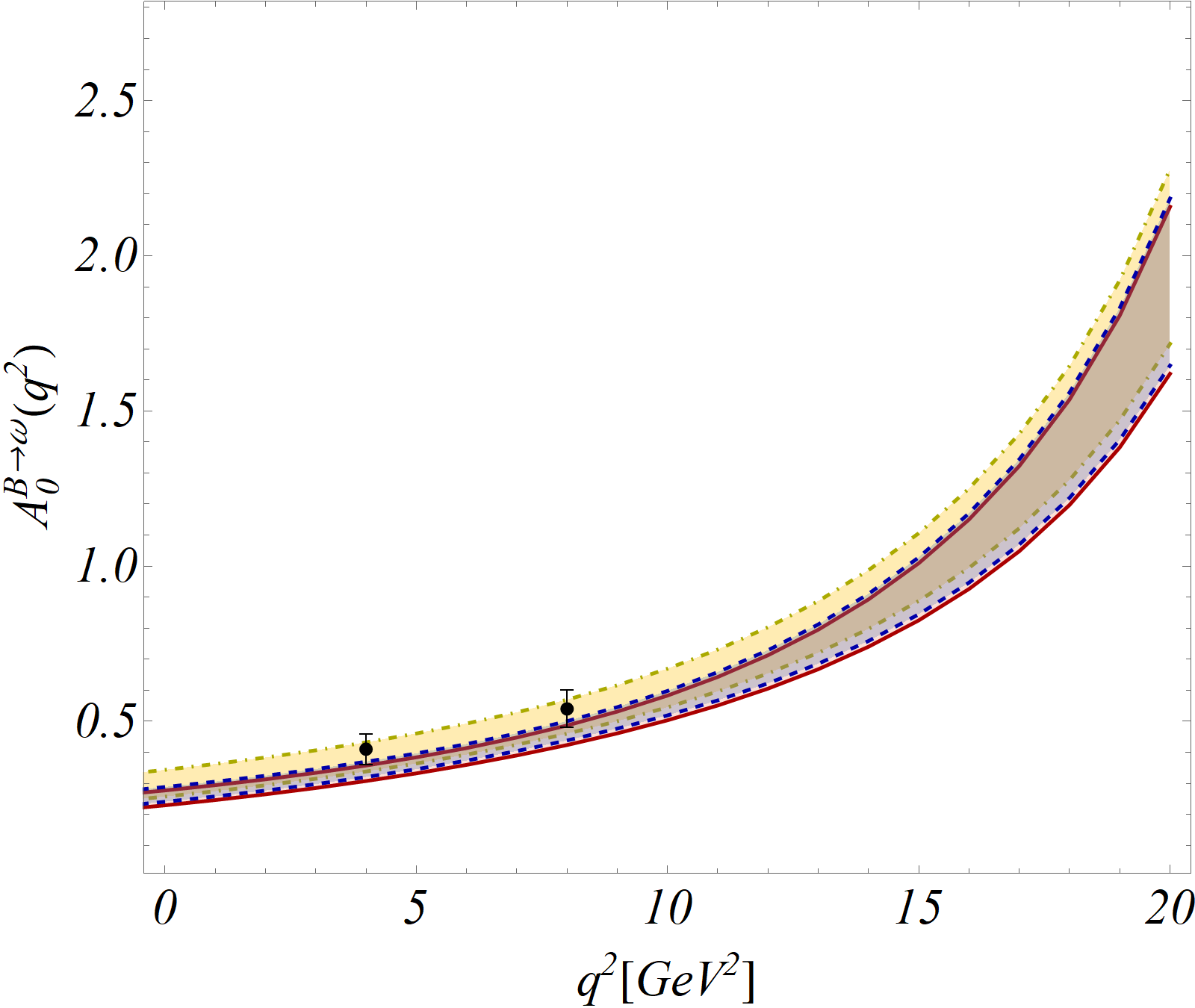}\label{fig:A0Omega}} ~~~~
  			\subfloat[]{\includegraphics[width=0.3\textwidth]{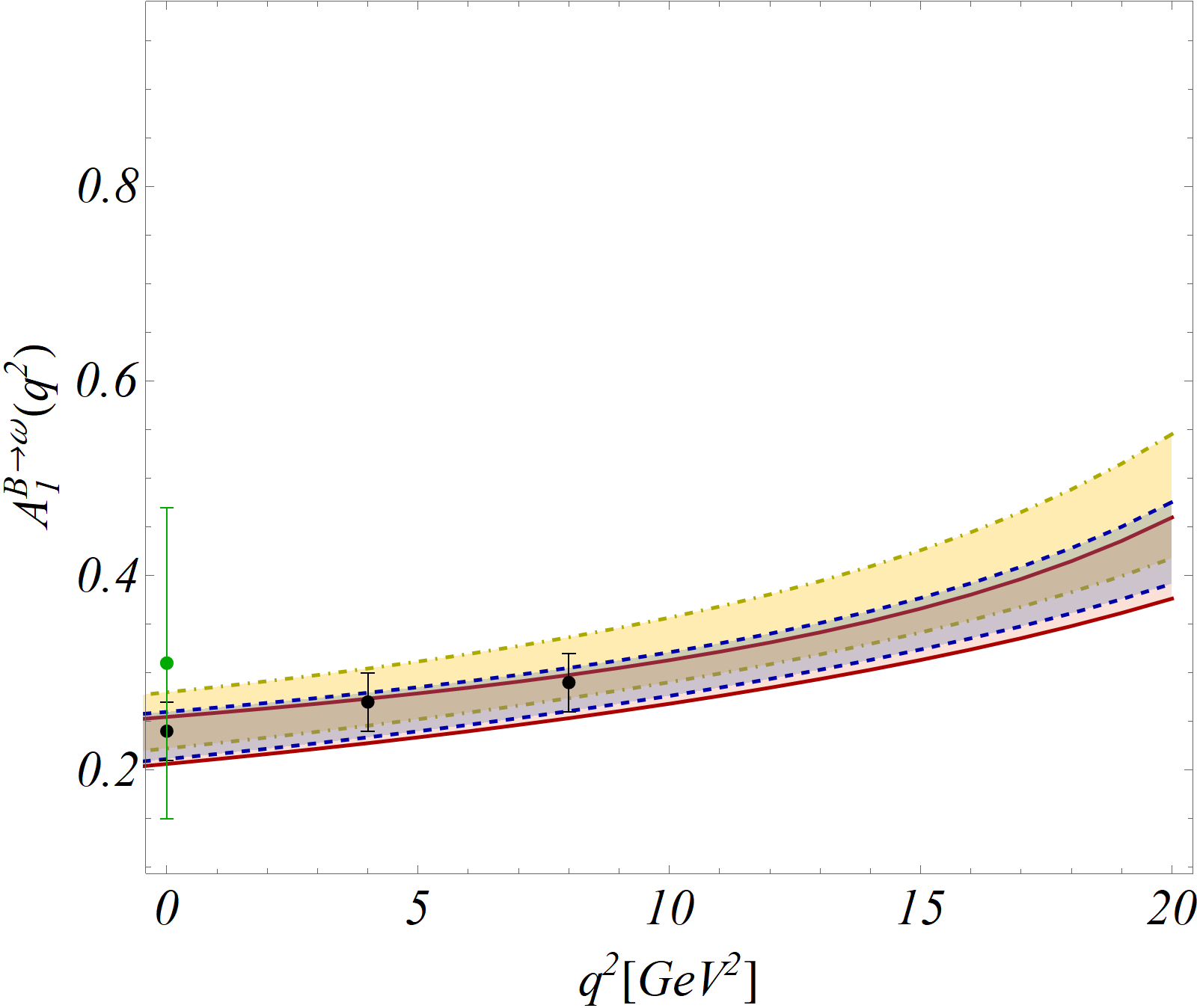}\label{fig:A1Omega}}~~~~	\subfloat[]{\includegraphics[width=0.3\textwidth]{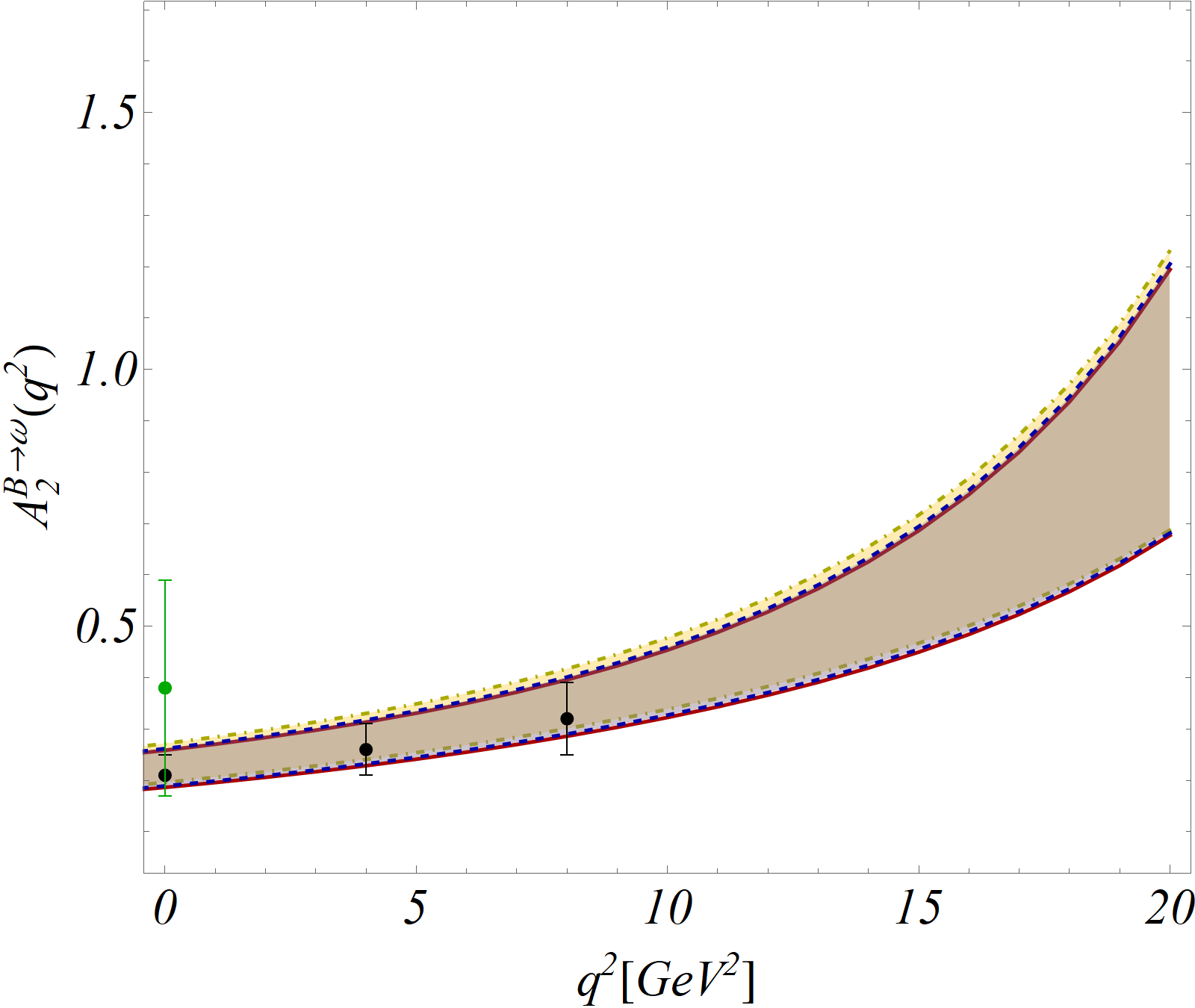}\label{fig:A2Omega}} \\
  			\subfloat[]{\includegraphics[width=0.3\textwidth]{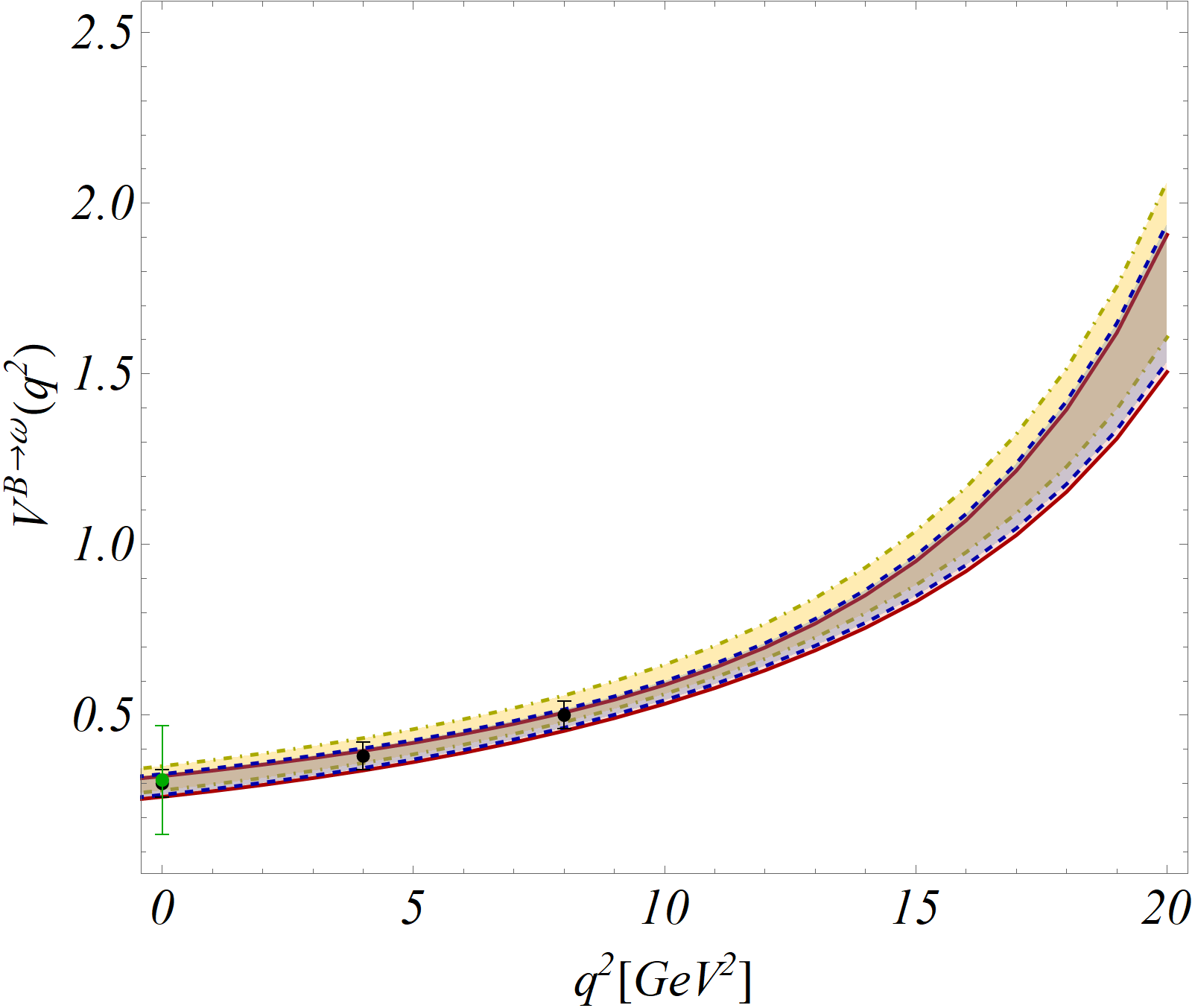}\label{fig:VOmega}}~~~~~~~
  			\subfloat[]{\includegraphics[width=0.25\textwidth]{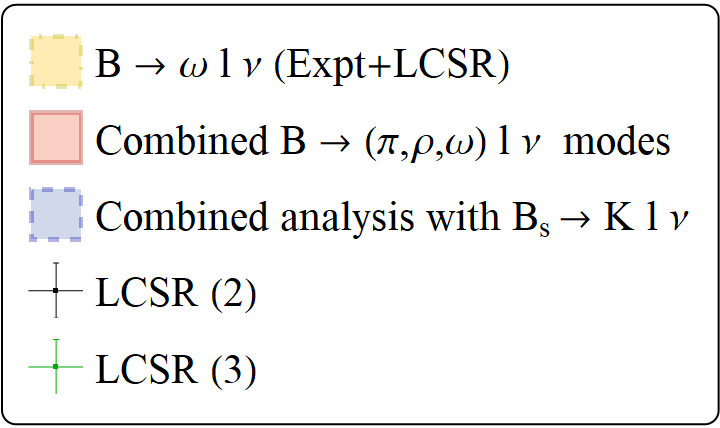}\label{fig:legOmega}}
  			\caption{The $q^2$ distributions of the form factors $A_0$, $A_1$, $A_2$ and $V$ for $\bar{B} \to \omega l^- \bar{\nu}$ modes. The LCSR datapoints are also shown in the plots. The labels ``LCSR(2)" and ``LCSR(3)" refer to the inputs from LCSR \cite{Straub:2015ica} and \cite{PhysRevD.101.074035} respectively.}
  			\label{fig:FFomega}
  		\end{figure*}

\subsection{$|V_{ub}|$ from $\bar{B} \to (\pi,\rho,\omega) \ell^- \bar{\nu}_{\ell}$ and $\bar{B}_s \to K \mu^- \bar{\nu}_{\mu}$ decays }
 
Data on the differential $\bar{B} \to \pi l^- \bar{\nu}$ decay rates is available from BaBar (2011)~\cite{delAmoSanchez:2010af}, Belle (2011)~\cite{Ha:2010rf}, BaBar (2012)~\cite{Lees:2012vv}, Belle (2013)~\cite{Sibidanov:2013rkk}. A $q^2$-averaged spectrum of all these measurements is obtained by HFLAV \cite{HFLAV:2019otj} which we have directly used in our analysis. The lattice collaborations RBC/UKQCD \cite{Flynn:2015mha} and JLQCD \cite{Colquhoun:2022atw} provide synthetic data points for $f_{+,0}(q^2)$ with full covariance matrices (both systematic and statistical) at three $q^2$ points which we have directly used in our analysis. On the other hand, the Fermilab/MILC collaboration \cite{FermilabLattice:2015cdh} provides the fit results for the BCL coefficients of the $z$-expansions of the respective form factors. We use their fit results to generate correlated synthetic data points at exactly the same $q^2$ values as RBC/UKQCD, with an extra point for $f_+$ at $q^2 = 20.5$ GeV$^2$. In addition to the lattice inputs, we have used the inputs on the form factors obtained by LCSR approaches \cite{Gubernari:2018wyi,Leljak:2021vte}. Note that due to the adoption of a different approach \footnote{The analysis in \cite{Leljak:2021vte} uses the two-particle twist-two pion light-cone distribution amplitude (LCDA). The results are more precise than those obtained in \cite{Gubernari:2018wyi}, which is an LO calculation with the ill-known B-meson LCDA.} the predictions in \cite{Gubernari:2018wyi} have large errors as compared to those given in ref.~\cite{Leljak:2021vte}. Ref. \cite{Leljak:2021vte} provides the form factors $f_+(q^2)$ at $q^2=-10,-5,0,5,10$ GeV$^2$ and $f_0(q^2)$ at $q^2=-10,-5,5,10$ GeV$^2$ respectively, whic play an essential role in our fits. Our results do not depend on the inputs from ref.~\cite{Gubernari:2018wyi}. The value of  $f_0$ at $q^2=0$ is obtained via the QCD relation $f_+(0) = f_0(0)$. The relevant details of these inputs and the correlations can be seen from our earlier publication \cite{Biswas:2022lhu}.

We have analysed all these available inputs in different possible combinations in the extractions of $|V_{ub}|$ from the $\bar{B} \to \pi l^- \bar{\nu}$ decays (Fit-(I) to Fit-(IX)) in table \ref{tab:Vub}. We have incorporated the HFLAV average for the available experimental data on these decays in all these fits. The impact of inputs from different lattice groups has been studied separately and compared. Also, in all these cases, we have checked the impact of the LCSR inputs \cite{Gubernari:2018wyi,Leljak:2021vte}. As we have mentioned earlier the fit is sensitive to the LCSR inputs from \cite{Leljak:2021vte} and practically insensitive to the inputs from \cite{Gubernari:2018wyi}. A couple of cases exhibit a shift in the best fit value of $|V_{ub}|$ after the inclusion of the LCSR inputs. Finally, we combine all the available information and extract $|V_{ub}|$ along with the form factor parameters. The results for $|V_{ub}|$ are given in the fifth column of table~\ref{tab:Vub}. Note that the extracted value of $|V_{ub}|$ is higher in the fit with the lattice data from JLQCD. However, it has a large error due to large uncertainties in the corresponding (lattice) data. The inclusion of LCSR in this fit reduces the error considerably along with a reduction of the best fit value by $\approx$ 6\%. The inclusion of LCSR also results in a similar observation for the fit with only RBC/UKQCD. In this case however, the best fit value of $|V_{ub}|$ is much lower than what we obtain in the fit with Fermilab/MILC or JLQCD alongside the LCSR. After combining all the inputs from the three different lattice groups along with all the experimental results on $\bar{B}\to \pi \ell^- \bar{\nu}_{\ell}$ decays, we obtain
\begin{equation}\label{eq:VubB2pi}
|V_{ub}| = (3.69 \pm 0.13)\times 10^{-3}.   
\end{equation}       
Note that the error is $\approx$ 3\%, which is due to the precise lattice inputs from Fermilab/MILC collaboration. As we can see from the table that the extracted value of $|V_{ub}|$ has $\approx$ 6\% error if we carry out the fit without the inputs from Fermilab/MILC. In the combined fit, the inclusion of LCSR does not change the result, since the inputs from Fermilab/MILC have played the dominant role due to their smaller errors than the other inputs in this scenario too. Note that in this combined analysis the new inputs are the lattice inputs from JLQCD. The predictions based on the data, Fermilab/MILC and RBC/UKQCD is given by $|V_{ub}| = (3.70 \pm 0.16)\times 10^{-3}$ \cite{HFLAV:2019otj,Biswas:2021qyq,pdg}.

 \begin{table}[t]
 	\begin{center}
 		\scriptsize
 		\renewcommand*{\arraystretch}{1.6}
 		\begin{tabular}{|c|c|c|c|c|c|}
 			\hline
 			Mode  & Fits scenarios of table \ref{tab:Vub} & $\chi^2_{min}$/DOF& p-value(\%) & $|V_{ub}|$ $\times$ $10^3$ & $|V_{ub}|/|V_{cb}|$\\
 			\hline
 			$\bar{B} \to \pi l^- \bar{\nu}$  & Fit-VIII  & 10.66/12 & 55.8 & 3.69 $\pm$ 0.20 &0.091 $\pm$ 0.005\\
 			& Fit-IX  & 15.4/30 & 98.7 & 3.67 $\pm$ 0.17 & 0.091 $\pm$ 0.004 \\
 			\hline 
 			$\bar{B} \to \pi l^- \bar{\nu}$ and & Fit-X & 11.7/13  & 54.9 & 3.57 $\pm$ 0.15 & 0.088 $\pm$ 0.004\\
 			$\bar{B}_s \to K l^- \bar{\nu}$ 
 			& Fit-XI & 29.8/32 & 57.7 & 3.41 $\pm$ 0.11 & 0.084 $\pm$ 0.003\\
 			\hline
 			$\bar{B} \to \pi(\rho,\omega) l^- \bar{\nu}$ &  All inputs combined & 58.6/88 & 99.3 & 3.51 $\pm$ 0.13 & 0.087 $\pm$ 0.003 \\			
 			\hline
 			&  $B\to \pi$ and $B_s \to K$:  & 53.9/71 & 93.5 & 3.46 $\pm$ 0.12 & 0.085 $\pm$ 0.003\\
 			& Experiment + Lattice & & & &\\
 			$\bar{B} \to \pi(\rho,\omega) l^- \bar{\nu}$ and		&   $B \to (\rho,\omega)$: Experiment   & & & &\\
 			&  + LCSR & & & &\\
 			\cline{2-6}
 			$\bar{B}_s \to K l^- \bar{\nu}$	  & For all modes: Experiment + & 70.7/90 & 93.4 & 3.36 $\pm$ 0.10 & 0.083 $\pm$ 0.003 \\
 			&  lattice + LCSR & & & &\\
 			\hline
 		\end{tabular}
 		\caption{Determination of $|V_{ub}|$ in various scenarios after adopting the procedure. }
 		\label{tab:Vubnew}
 	\end{center}
 \end{table} 

In Fig.~\ref{fig:FFpi} we have plotted the $q^2$ dependence of $f_{+,0}^{B\to\pi}(q^2)$ obtained from the above fit results to $\bar{B}\to \pi\ell^- \bar{\nu}$ decays. We have compared the shapes obtained from Fit-VIII and Fit-IX, which shows that including LCSR inputs does not change the shapes. We have also plotted the shapes obtained from Fit-(I), which have large errors but are consistent with the ones obtained from the other two fits. Note that in all the fits the shapes for $f_+^{B\to\pi}(q^2)$ are very precise and consistent with the lattice and the respective LCSR inputs. However, the fit values for $f_0^{B\to\pi}$ at $q^2=5$ and 10 GeV$^2$ are inconsistent with the respective LCSR inputs \cite{Leljak:2021vte} at their 1-$\sigma$ ranges. The fit value for $f_0^{B\to\pi}(q^2=0)$ is fully consistent with the LCSR results from the fits with and without the LCSR inputs. Furthermore, the lattice input from Fermilab/MILC at $q^2=19$ GeV$^2$ is slightly off as compared to the shapes obtained from Fit-VIII and Fit-IX.  
The inputs on the form factors corresponding to $B_s \to K$ transitions are available from Lattice QCD collaborations like Fermilab/MILC \cite{PhysRevD.100.034501}, RBC/UKQCD \cite{Flynn:2015mha} and HPQCD \cite{HPQCD:BStoK2014}. RBC/UKQCD provide the pseudo data points at $q^2 = 17.6$, $20.8$, and $23.4$ GeV$^2$ respectively, which we have directly used in our analyses. Also, we have generated synthetic lattice data points at $q^2 = 17.6$, $20.8$, and $23.4$ GeV$^2$ respectively, from the fit results of the form factor parameters given in the Fermilab/MILC and HPQCD analyses. Ref.~\cite{Khodjamirian:2017fxg} reports the value of the form factor $f_+^{B_s \to K}$ at $q^2 = 0$ only following the LCSR approach. In the BSZ expansion (eq.~\ref{eq:bszexp}) for the form factors in $B_s \to K$ decays, we have considered terms up to $N=2$. In table \ref{tab:Vub} (Fit-X to Fit-XIII), we have presented the extracted value of $|V_{ub}|$ from the combined analyses of the available information on $\bar{B}\to \pi\ell^- \bar{\nu}$ and $\bar{B}_s\to K\mu^- \bar{\nu}$ decays. In addition to all the available experimental data and theory inputs on $\bar{B}\to \pi\ell^- \bar{\nu}$ decays as discussed above, we include the data on $\mathcal{B}(\bar{B}_s\to K^+\mu^- \bar{\nu}_{\mu})$ (eq.~\ref{eq:totalBstok}), the lattice and LCSR inputs on $f_{+,0}^{B_s\to K}(q^2) $. In addition, here we carry out a few more analyses to see the impact of different theory inputs. From the results obtained, we note that the extracted values of $|V_{ub}|$ from these combined fits are sensitive to the lattice inputs from HPQCD as well as to $f_{+}^{B_s\to K}(0) $ from LCSR \cite{Khodjamirian:2017fxg}. Though in all these fits, the extracted values of $|V_{ub}|$ are consistent with each other, the best fit value has increased by $\approx$ 4\% after dropping the inputs from HPQCD. 
Including all the lattice and experimental data in the fit, we have obtained 
\begin{equation}\label{eq:VubB2piBs2K}
|V_{ub}| = (3.60 \pm 0.11)\times 10^{-3}. 
\end{equation}  
The exclusion of the lattice inputs on $f_{+,0}^{B_s \to K}$ from HPQCD results in a 4\% increase in the best-fit point of the extracted value of $|V_{ub}|$.

In Fig.~\ref{fig:FFBsK}, we have compared the $q^2$ variation of $f_{+,0}^{B_s\to K}(q^2)$ obtained from the results of the combined fits: Fit-X, Fit-XI, Fit-XII given in table \ref{tab:Vub}. In the same plots, we have also shown the $q^2$ shapes of $f_{+,0}^{B_s\to K}(q^2)$ obtained only from the lattice inputs on $B_s\to K$ form factors. Note that the $f_{+,0}^{B_s\to K}(q^2)$ shapes obtained only from the lattice data are consistent with those obtained from Fit-X and Fit-XI, respectively. However, the $q^2$ shapes for $f_{+,0}^{B_s\to K}$ obtained from Fit-XII, where the HPQCD inputs for $f_{+,0}^{B_s\to K}(q^2)$ have been dropped, are slightly inconsistent with the rest of the three distributions within 1-$\sigma$ in the low-$q^2$ regions. All of them are pretty consistent at high $q^2$ though. In table \ref{tab:binpred1} in the appendix we have predicted the branching fractions $\mathcal{BR}(\bar{B}_s \to K \ell^-\bar{\nu}_{\ell})$ in different $q^2$-bins using a few fit results given in table \ref{tab:Vub}.    

\begin{table}[t]
	\begin{center}
		\small
		\begin{tabular}{|c|c|c|c|}
			\hline
			\multicolumn{4}{|c|}{Experimental : $R_{BF}$ (Low) = (1.66 $\pm$ 0.12) $\times$ $10^{-3}$ , $R_{BF}$ (High) = (3.25 $\pm$ 0.28) $\times$ $10^{-3}$ }\\
			\hline
			Analysed Mode(s) & Scenario (Inputs) & \multicolumn{2}{c|} {Predictions of $R_{BF}$ ($10^{-3}$)} \\
			
			\cline{3-4}
			& & Low ($q^2$ $<$ 7 $\text{GeV}^2$) & High ($q^2$ $>$ 7 $\text{GeV}^2$) \\
			\hline
			$\bar{B} \to \pi \ell^- \bar{\nu}_{\ell}$, $b \to c \ell^- \bar{\nu}_{\ell}$, and & Experiment + Lattice  & 1.98 $\pm$ 0.42 & 3.79 $\pm$ 0.53 \\
			no experimental inputs  &   & &  \\
			on $\mathcal{BR}(\bar{B}_s\to K\mu^- \bar{\nu}_{\mu})$	& Experiment+Lattice+LCSR & 2.59 $\pm$ 0.43 & 4.14 $\pm$ 0.51 \\
			\hline
			$\bar{B} \to \pi \ell^- \bar{\nu}_{\ell}$, $b \to c \ell^- \bar{\nu}_{\ell}$, and &Experiment+Lattice &  1.65 $\pm$ 0.21 & 3.38 $\pm$ 0.26\\
			$\mathcal{BR}(\bar{B}_s\to K\mu^- \bar{\nu}_{\mu})$ & &   & \\
			& Experiment+Lattice+LCSR & 2.10 $\pm$ 0.19 & 3.37 $\pm$ 0.24\\
			\hline
			$\bar{B} \to (\pi,\rho,\omega) \ell^- \bar{\nu}_{\ell}$, $b \to c \ell^- \bar{\nu}_{\ell}$,  & & & \\
			and no experimental inputs & Experiment+Lattice+LCSR & 2.37 $\pm$ 0.37 & 3.78 $\pm$ 0.42 \\
			on $\mathcal{BR}(\bar{B}_s\to K\mu^- \bar{\nu}_{\mu})$ & &  & \\	
			\hline
			All $b \to c(u) l^- \bar{\nu}$ & Experiment+Lattice+LCSR & 2.08 $\pm$ 0.18 & 3.31 $\pm$ 0.23\\
			\hline
		\end{tabular}
		\caption{Predictions for the ratios of the branching fractions obtained in the different fit scenarios. Here, $R_{BF}$ refers to $\frac{\mathcal{BR}(\bar{B}_s \to K \mu^- \bar{\nu})}{\mathcal{BR}(\bar{B}_s \to D_s \mu^- \bar{\nu})}$.}
		\label{tab:pred1}
	\end{center}
\end{table}

Our analyses also include the available inputs on $\bar{B} \to \rho(\omega) l^- \bar{\nu}$ decays. For $\bar{B} \to \rho(\omega) l^- \bar{\nu}$ transitions, the experimental inputs for these channels correspond to the measurements on the partial branching fractions in bins of $q^2$ from BaBar, and Belle collaborations \cite{delAmoSanchez:2010af,BaBar:2012dvs,Lees:2012vv,Sibidanov:2013rkk}\footnote{Note that in ref.~\cite{Bernlochner:2021rel}, the average of the $q^2$ spectra measured for the $\bar{B}\to\rho \ell^-\bar{\nu}_{\ell}$ and $\bar{B}\to\omega \ell^-\bar{\nu}_{\ell}$ decays at the Belle and BaBar have been generated. For the average in $\bar{B}\to \rho\ell^-\bar{\nu}_{\ell}$, the analysis uses two uncorrelated datasets from Belle and BaBar. Also, for $\bar{B}\to \omega \ell^-\bar{\nu}_{\ell}$ the analysis in \cite{Bernlochner:2021rel} does not include the data from BaBar 2012 analysis \cite{Lees:2012vv}. They have only used the datasets from BaBar and Belle, which are uncorrelated. Therefore, we preferred to analyse all the independent measurements as separate inputs.}. The situation is a bit grim for the form factors associated with the $\bar{B}\to\rho$ and $\bar{B}\to\omega$ transitions. Both the $\bar{B} \to \rho\ell^- \bar{\nu}_{\ell}$ and $\bar{B} \to \omega\ell^- \bar{\nu}_{\ell}$ decays are sensitive to the four form factors: $V(q^2)$, $A_0(q^2)$, $A_1(q^2)$ and $A_2(q^2)$ for which no lattice estimates exist till date. All these form factors are estimated in the LCSR approach, which is available for a few values of $q^2$, including the one at $q^2=0$ \cite{PhysRevD.101.074035,Straub:2015ica,Gubernari:2018wyi}. All these form factors are fully correlated. For $\bar{B}\to\rho$ transitions, the LCSR data points along with the respective covariance matrices are given in ref.~\cite{Gubernari:2018wyi} for $q^2=-15, -10,-5,0, 5$ GeV$^2$, which we directly use in our fits. In ref.~\cite{Straub:2015ica}, the fit results for the coefficients of the $z$-expansion are given using which we generate correlated synthetic data points for the form factors $A_1$, $A_2$ and $V$ at $q^2$ = 0,5,10 GeV$^2$ and for $A_0$ at $q^2$ = 5,10 GeV$^2$ \footnote{At $q^2$ = 0, $A_0$ is related to $A_1$ and $A_2$, thus it isn't independent. So, in order to keep the covariance matrix positive semi-definite, we don't include the datapoint for $A_0$ at $q^2$ = 0.} respectively, for the $\bar{B}\to \rho$ transitions. For $\bar{B}\to\omega$ transitions, we generate correlated synthetic data points for the form factors $A_1$, $A_2$ and $V$ at $q^2$ = 0,4,8 GeV$^2$ and for $A_0$ at $q^2$ = 4,8 GeV$^2$. Also, here for both the modes for the BSZ expansion of the form factors, we have truncated the series at $N=2$, which can be improved further once the inputs from the lattice are available. For both these decays, from the analyses of the independent modes, we obtain relatively lower values of $|V_{ub}|$ though they have large errors and are consistent with the ones from eqs.~\ref{eq:VubB2pi} and \ref{eq:VubB2piBs2K} respectively. Hence the combination of both these modes with the $\bar{B}\to \pi\ell^- \bar{\nu}$ and $\bar{B}_s\to K \mu^- \bar{\nu}_{\mu}$ decays will result in a lower value of $|V_{ub}|$ than that expected from the combined or independent analyses involving the $\bar{B}\to \pi\ell^- \bar{\nu}$ and $\bar{B}_s\to K \mu^- \bar{\nu}_{\mu}$ decays. 

In Fig.~\ref{fig:FFrho}, we determine and compare the $q^2$ shapes of the form factors in $\bar{B}\to \rho$ decays. We extract the shapes from the different fits corresponding to the independent modes as well as from the combined analyses with and without the inputs from $\bar{B}_s\to K\mu^- \bar{\nu}_{\mu}$ decays. Note that the shapes obtained from the analysis of only $\bar{B}\to \rho\ell^- \bar{\nu}$ decay modes are consistent with those obtained from the two other analyses (combined analyses) in the last two rows of table \ref{tab:Vub}. The shapes are also consistent with the LCSR inputs used in the analyses. We observe a similar pattern in the extracted shapes of the form factors shown in Fig.~\ref{fig:FFomega} for $\bar{B}\to \omega$ decays. To further improve the results from the combined analysis, we need to wait for the lattice results of these vector modes and more precise measurements. The predictions for the branching fractions $\mathcal{BR}(\bar{B}\to \rho\ell^- \bar{\nu}_{\ell})$ and $\mathcal{BR}(\bar{B}\to \omega\ell^- \bar{\nu}_{\ell})$ in a few small $q^2$ bins are given in table \ref{tab:binpred} in the appendix, along with a follow-up discussion.

In the last column of table \ref{tab:Vub}, we have shown the extracted values of the ratio $|V_{ub}|/|V_{cb}|$ for each of the fit scenarios. In all these estimates, the value of $|V_{cb}|$ has been taken from the ``all modes" fit in table \ref{tab:Vcb} (second last row). Note that all these estimates are more or less consistent with the measured value from the ratio of the baryonic decay rates given in eq.~\ref{eq:VubVcbbaryon}. 

Note that the fit to data and all the available lattice inputs in $\bar{B}\to \pi\ell^- \bar{\nu}$ and $\bar{B}_s\to K\mu^- \bar{\nu}$ decays, separately and combinedly, have $\chi^2/{d.o.f}$ larger than one. For example, the fits (VIII) and (X) in table \ref{tab:Vub} have $\chi^2/{d.o.f} \approx 2.12$ and $1.97$, respectively. This is due to the precise Lattice inputs from the Fermilab/MILC collaboration. Following the method of PDG, which has also been adopted by FLAG \cite{FlavourLatticeAveragingGroupFLAG:2021npn}, one can rescale the error of the fitted parameters by the factor $\sqrt{\chi^2/{d.o.f}}$ which does not have any impact on the respective correlations. This method will increase the error of the fitted parameters. In this paper, we have used this rescaling method slightly differently though the extracted error will be the same as one can obtain from a direct rescaling by $\sqrt{\chi^2/{d.o.f}}$ given in table \ref{tab:Vub}. We have noticed that the combined fit to the z-expansion coefficients with all the Lattice inputs for the $\bar{B} \to \pi$ mode has $\chi^2/d.o.f$ = 40.0/12 and for the $\bar{B}_s \to K$ mode has $\chi^2/d.o.f$ = 22.5/13. We then rescale the uncertainties of the $z$-expansion coefficients in $B \to \pi$ and $B_s \to K$ form factors by the factors $\sqrt{40.0/12} = 1.83$ and $\sqrt{22.5/13} = 1.32$, respectively. We use these form-factor parameters as nuisance parameters in the subsequent fits with the experimental data and LCSR inputs. The results are shown in table \ref{tab:Vubnew}. We find that the p-values of the fits have increased, and also, the uncertainties on $|V_{ub}|$ have increased as compared to the results of table \ref{tab:Vub}. We have adopted these results for the $z$-expansion coefficients in the rest of our analysis, as discussed below.

\begin{table}[t]
	\begin{center}
		\scriptsize
		\renewcommand*{\arraystretch}{1.4}
		\begin{tabular}{|c|c|c|c|c|c|c|}
			\hline
			Mode(s) & Inputs & $\chi^2_{min}$/DOF  & $|V_{ub}|$ $\times$ $10^3$ & $|V_{cb}|$ $\times$ $10^3$ &  $\frac{|V_{ub}|}{|V_{cb}|}$ & Correlation $(\%)$ \\
			 &  &  &  & & & [$|V_{ub}|, |V_{cb}|$] \\
			\hline
			$\bar{B} \to \pi l^- \bar{\nu}$   & Experiment+ All Lattice & 171.1/156 & 3.61 $\pm$ 0.15 & 40.4 $\pm$ 0.6 &   0.089 $\pm$ 0.004 & 12.8 \\
		  and   &  Experiment+All Lattice+LCSR & 191.7/167  & 3.43 $\pm$ 0.12 & 40.4 $\pm$ 0.6 & 0.085 $\pm$ 0.003 & 14.1 \\
	        \cline{2-7}
		$b \to c l^- \bar{\nu}$	& Experiment+ All Lattice &  176.1/166 & 3.57 $\pm$ 0.13 & 40.3 $\pm$ 0.6 & 0.089 $\pm$ 0.003 & 11.5 \\
			& +LCSR (without $f_+^{B_s \to K}$) & & & & & \\ 
			\hline
			$b \to c l^- \bar{\nu}$, & Experiment+Lattice+LCSR  & 231.8/225 & 3.34 $\pm$ 0.10 & 40.4 $\pm$ 0.6 & 0.083 $\pm$ 0.003 & 12.4 \\
			 \cline{2-7}
				$b \to u l^- \bar{\nu}$	&  Experiment+ All Lattice  & 219.5/224 & 3.51 $\pm$ 0.11 & 40.2 $\pm$ 0.6 & 0.087 $\pm$ 0.003 & 10.4 \\
			 & +LCSR (without $f_+^{B_s \to K}$) & & & & & \\
			\hline
		\end{tabular}
		\caption{Extractions of the ratio $|V_{ub}|/|V_{cb}|$ from the fits to $b \to c(u) l \nu$ modes, including the ratio $R_{BF}$ = $\frac{\text{BR}(B_s \to K \mu \nu)}{\text{BR}(B_s \to D_s \mu \nu)}$ in the two bins of $B_s \to K$ momentum transfer \cite{LHCb:2020ist}. }
		\label{tab:ratio}
	\end{center}
\end{table}

\begin{figure*}[t]
	\small
	\centering
	\subfloat[]{\includegraphics[width=0.40\textwidth]{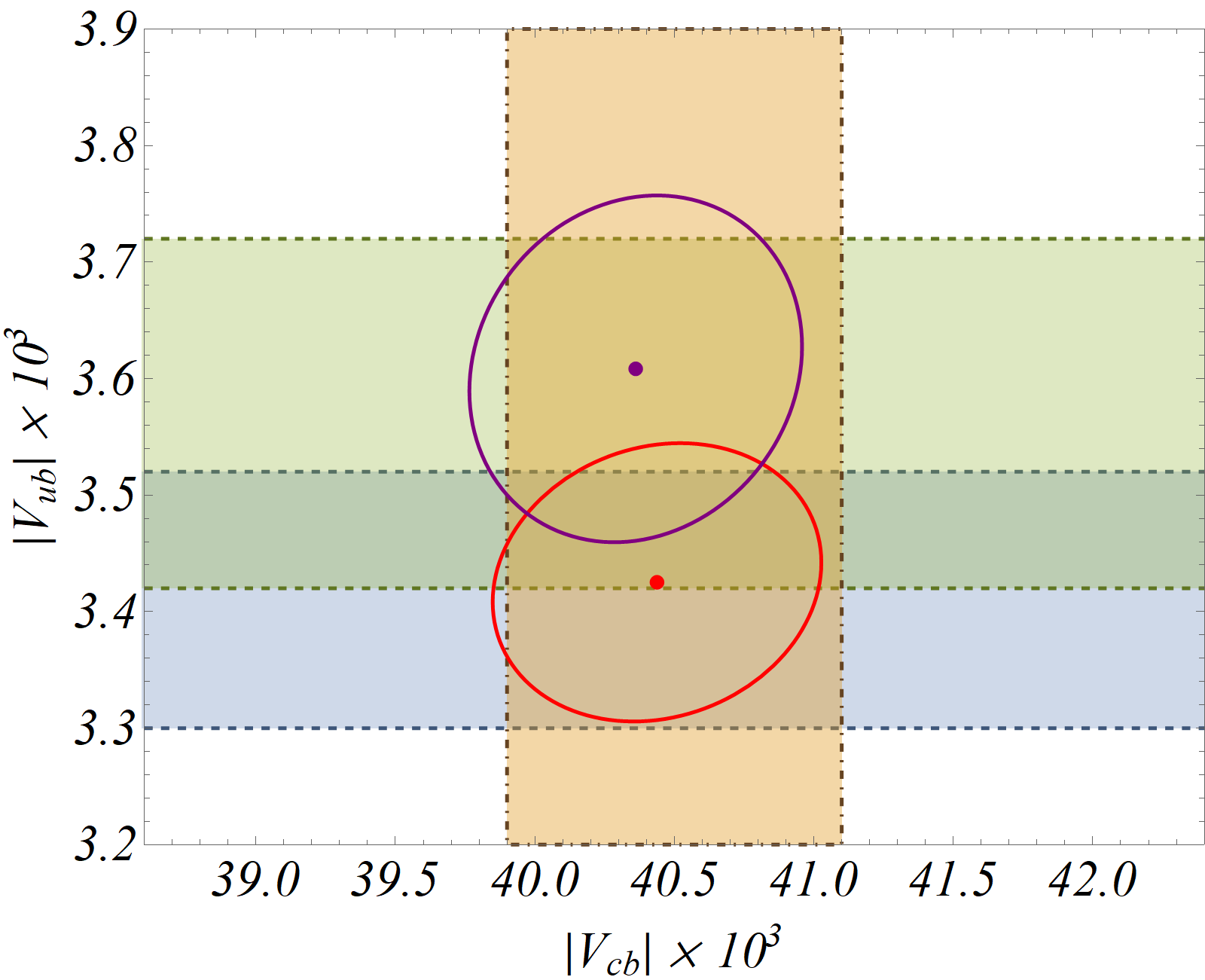}\label{fig:corrplt}}~~~~~
	\subfloat[]{\includegraphics[width=0.33\textwidth]{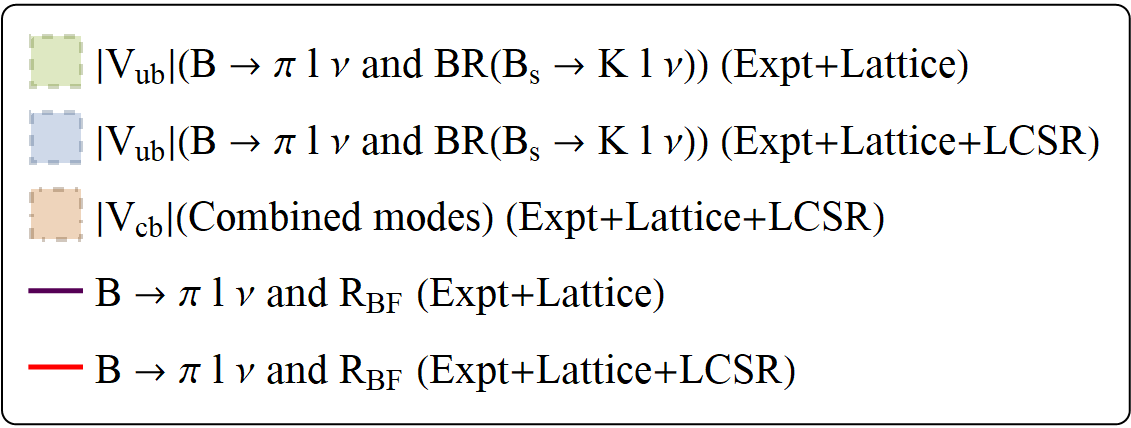}\label{fig:legcorr}}
	\caption{1$\sigma$ allowed regions and the respective correlations in the $|V_{ub}|$ and $|V_{cb}|$ plane. The legend shows the scenarios with the corrresponding colour coding. The respective numbers are taken from the table~\ref{tab:Vcb}, and the specific cases in tables~\ref{tab:Vub} and \ref{tab:ratio}, respectively. }
	\label{fig:corr}
\end{figure*}

Furthermore, for the cases discussed in table \ref{tab:Vub}, we have predicted the ratio of the branching fractions defined in eq.~\ref{eq:ratio} in the low and high-$q^2$ regions. In such predictions, we need the value of $|V_{ub}|^2/|V_{cb}|^2$, for which we consider the extracted values provided in table \ref{tab:Vubnew}. In table \ref{tab:pred1}, we have presented the results for a few specific cases of the combined analyses. We divide the results into two main segments depending on whether or not the experimental input on $\mathcal{BR}(\bar{B}_s\to K\mu^- \bar{\nu}_{\mu})$ has been considered in the fits. Note that in both segments, we need to consider lattice/LCSR inputs on $f_{+,0}^{B_s\to K}$ to get the predictions for the ratios. For the fit scenario without the input $\mathcal{B}(\bar{B}_s\to K\mu^- \bar{\nu}_{\mu})$, the predictions for the ratios  are given by 
\begin{align}\label{eq:RBFSM1}
R_{BF}^{low}|_{SM-1} &= (1.98 \pm 0.42 )\times 10^{-3}, \nn \\
R_{BF}^{high}|_{SM-1} &= (3.79 \pm 0.53 )\times 10^{-3}, 
\end{align}
In these predictions, the shape of the form factors $f_{+,0}^{B_s\to K}(q^2)$ depend only on the lattice inputs for these form factors, not on the data on $\bar{B}_s \to K\mu^- \bar{\nu}_{\mu}$. In the alternative scenario where the extraction of $|V_{ub}|/|V_{cb}|$ depends on the experimental result for $\mathcal{BR}(\bar{B}_s\to K\mu^- \bar{\nu}_{\mu})$, the predictions are given by 
 \begin{align}
 R_{BF}^{low}|_{SM-2} &= (1.65 \pm 0.21 )\times 10^{-3}, \nn \\
 R_{BF}^{high}|_{SM-2} &= (3.38 \pm 0.26 )\times 10^{-3}.
 \end{align}
These values are slightly less than the ones given in eq.~\ref{eq:RBFSM1}, which is due to a smaller value of $|V_{ub}|/|V_{cb}|$ obtained in the corresponding case (see table \ref{tab:Vubnew}). In all these scenarios, the inclusion of the LCSR inputs on $f_{+}^{B_s\to K}(q^2=0) = f_{0}^{B_s\to K}(q^2=0)$ increases the respective predicted values of $R_{BF}$ in both the $q^2$ regions. It is even seen that the predictions with the LCSR inputs are not consistent with the individual experimental results for the scenarios when $\mathcal{B}(\bar{B}_s\to K\mu^- \bar{\nu}_{\mu})$ are not included in the fits.

Finally, we carry out a couple of fits using the experimental results on $R_{BF}^{low}$ and $R_{BF}^{high}$ (eqs.~\ref{eq:ratio}) alongside the other inputs discussed earlier other than $\mathcal{BR}(\bar{B}_s\to K\mu^- \bar{\nu}_{\mu})$. Like before, we have checked the impact of different inputs on the results. The different fits and the corresponding results with the respective errors are shown in table \ref{tab:ratio}. For all the fits, we have extracted $|V_{ub}|$, $|V_{cb}|$, and their respective correlation, using which we have estimated the ratio $|V_{ub}|/|V_{cb}|$. In all the different fits, the extracted values of the ratio of the CKM elements are consistent. They also agree with the measured value from the baryonic decays given in eq.~\ref{eq:VubVcbbaryon}. As was observed earlier, the inclusion of LCSR input on $f_+^{B_s\to K}(q^2=0)$ reduces the value of the ratio and $|V_{ub}|$ in these fits too. Furthermore, the estimated values of $|V_{cb}|$ agree with the results given in eq.~\ref{eq:Vcbpred}. Note that in these fits, there will be a correlation between $|V_{ub}|$ and $|V_{cb}|$, which are shown in the last column in table \ref{tab:ratio}. In Fig.~\ref{fig:corrplt}, we have shown the allowed regions in the $|V_{ub}|-|V_{cb}|$ plane for a few specific fits including the $\bar{B}\to \pi\ell^- \bar{\nu}_{\ell}$, $\bar{B}_s\to K\mu^- \bar{\nu}_{\mu}$ and all $b\to c\ell^- \bar{\nu}_{\ell}$ modes. We have also shown the correlations between $|V_{ub}|$ and $|V_{cb}|$ from the fits with $R_{BF}$ instead of $\mathcal{BR}(\bar{B}_s\to K\mu^- \bar{\nu}_{\mu})$. Corresponding to each fit scenario defined in table \ref{tab:ratio}, we have estimated the ratio $R_{BF}$ in the low and high $q^2$ regions. The numbers are presented in table \ref{tab:pred} in the appendix along with the respective errors in their 1-$\sigma$ CI.

\section{Summary}

In this paper, we have extracted the CKM element $|V_{ub}|$ from the combined studies of the $\bar{B}\to (\pi,\rho,\omega)\ell^- \bar{\nu}_{\ell}$ and $\bar{B}_s\to K\mu^- \bar{\nu}_{\mu}$ decays, and $|V_{cb}|$ from a combined study of $\bar{B}\to D^{(*)}\ell^-\bar{\nu}_{\ell}$ and $\bar{B}_s\to D_s^{(*)}\ell^-\bar{\nu}_{\ell}$ decays. We have divided our analysis into several scenarios to see the impact of different inputs. For each of these scenarios, we have extracted the ratio of the CKM elements $\frac{|V_{ub}|}{|V_{cb}|}$, and the ratio of the decay rates $R_{BF} = \frac{\Gamma(\bar{B}_s \to K\mu^- \bar{\nu}_{\mu})}{\Gamma(\bar{B}_s \to D_s\mu^-\bar{\nu}_{\mu})}$. Also, for a consistency check, we have separated our analyses into two segments based on whether or not the measured value $\mathcal{BR}(\bar{B}_s \to K\mu^- \bar{\nu}_{\mu})$ has been considered as an input. In both types of cases, we have found a good agreement between the measured value of $\frac{|V_{ub}|}{|V_{cb}|}$ obtained by LHCb from the measurement of $\frac{\Gamma(\Lambda_b \to p \mu^-\bar{\nu}_{\mu})}{\Gamma(\Lambda_b \to \Lambda_c\mu^- \bar{\nu}_{\mu})}$ and our estimated values from the semileptonic mesonic decays. From the combined analysis we obtain $|V_{cb}| = (40.5\pm 0.6)\times 10^{-3}$. The following are our best results for $|V_{ub}|$: 
$$
|V_{ub}| = 
\begin{cases}
 (3.69 \pm 0.20)\times 10^{-3},\ \  \text{From $\bar{B}\to \pi\ell^-\bar{\nu}_{\ell}$, analysing the data and lattice,}\\
 (3.57 \pm 0.15)\times 10^{-3},\ \  \text{From $\bar{B}\to \pi\ell^-\bar{\nu}_{\ell}$ and $\bar{B}_s\to K\mu^-\bar{\nu}_{\mu}$, analysing the data and lattice,}\\
 (3.51 \pm 0.13)\times 10^{-3},\ \  \text{From $\bar{B}\to (\pi,\omega,\rho) \ell^- \bar{\nu}_{\ell}$, analysing the data, lattice and LCSR,}\\
 (3.46 \pm 0.12)\times 10^{-3},\ \  \text{From $\bar{B}\to (\pi,\omega,\rho) \ell^- \bar{\nu}_{\ell}$ and $\bar{B}_s\to K\mu^-\bar{\nu}_{\mu}$, analysing the data, lattice and LCSR.}\\
\end{cases}
$$
Using the above results, we respectively obtain:
$$
\frac{|V_{ub}|}{|V_{cb}|} = 
\begin{cases}
0.091 \pm 0.005,\ \  \text{From $\bar{B}\to \pi\ell^-\bar{\nu}_{\ell}$ and all $b\to c\ell^-\bar{\nu}_{\ell}$ modes, }\\
0.088 \pm 0.004 ,\ \  \text{From $\bar{B}\to \pi\ell^-\bar{\nu}_{\ell}$, $\bar{B}_s\to K\mu^-\bar{\nu}_{\mu}$, and all $b\to c\ell^-\bar{\nu}_{\ell}$ modes,}\\
0.087 \pm 0.003 ,\ \  \text{From $\bar{B}\to (\pi,\omega,\rho) \ell^- \bar{\nu}_{\ell}$, and all $b\to c\ell^-\bar{\nu}_{\ell}$ modes,}\\
0.085 \pm 0. 003 ,\ \  \text{From $\bar{B}\to (\pi,\omega,\rho) \ell^-\bar{\nu}_{\ell}$, $\bar{B}_s\to K\mu^-\bar{\nu}_{\mu}$, and all $b\to c\ell^-\bar{\nu}_{\ell}$ modes.}\\
\end{cases}
$$
Our estimated values of $\frac{|V_{ub}|}{|V_{cb}|}$ with and without the input from $\bar{B}_s\to K\mu^-\bar{\nu}_{\mu}$ agree with that measured by LHCb from the data on $R_{BF}$ in the high-$q^2$ regions but not with the one obtained in the low-$q^2$ regions. In addition, our estimated values of $R_{BF}$ in the low and high-$q^2$ regions from the analyses with $\mathcal{BR}(\bar{B}_s \to K\mu^-\bar{\nu}_{\mu})$ as input alongside all the other inputs more or less agree with the respective measured values. Without the inclusion of $\mathcal{BR}(\bar{B}_s \to K\mu^-\bar{\nu}_{\mu})$ in the fit, the estimated values of $R_{BF}$ disagree with the respective measured values when we include $f_+^{B_s\to K}(q^2=0)$ in the fit. We have shown the $q^2$ shapes of the form factors for each of the decays considered. In addition, using the results of fits, we have given the SM predictions for the branching fractions $\mathcal{BR}(\bar{B}\to \rho\ell^-\bar{\nu}_{\ell})$, $\mathcal{BR}(\bar{B}\to \omega\ell^-\bar{\nu}_{\ell})$, and $\mathcal{BR}(\bar{B}_s\to K\ell^-\bar{\nu}_{\ell})$ in small $q^2$ bins. In most bins, the currently available inputs allow predictions with an accuracy $\lsim 10$\%. However, in a few bins, the error could be larger than 10\%. We have to wait for more precise inputs from lattice and experiments for further improvements.              

We have also done the fits including the measured values of $R_{BF}$ instead of $\mathcal{BR}(\bar{B}_s \to K\mu^-\bar{\nu}_{\mu})$ along with the other inputs and extracted the ratio $\frac{|V_{ub}|}{|V_{cb}|}$ which are in good agreement with that measured from  
$\frac{\Gamma(\Lambda_b \to p \mu^-\bar{\nu}_{\mu})}{\Gamma(\Lambda_b \to \Lambda_c\mu^- \bar{\nu}_{\mu})}$. For all these different analyses, using the fit results, we are able to reproduce the measured values of $R_{BF}$ in both the low and high-$q^2$ regions with reasonably good consistency.

\acknowledgments
We would like to thank Marcello Rotondo and Mirco Dorigo for some useful communication. Also, we would like to thank Paolo Gambino for pointing out an error in our first version of the draft.  
This work of SN is supported by the Science and Engineering Research Board, Govt. of India, under the grant CRG/2018/001260.

\section{Appendix}

For completeness, in table \ref{tab:binpred1}, we provide the predictions for the $\mathcal{BR}(\bar{B}_s\to K \mu^- \bar{\nu}_{\mu})$ in the SM for a few small $q^2$ bins which can be tested in future experiments. We have considered inputs from lattice for the shape of the form factors and also checked the results, including the LCSR input $f_{+,0}^{B_s\to K}(q^2=0)$. Hence, the predictions are solely based on lattice or ``lattice + LCSR". For these predictions, we have considered $|V_{ub}|$ from Fit-VIII or Fit-IX given in table \ref{tab:Vub}. The removal of HPQCD inputs for the form factors reduces the predicted values of the branching fractions in the low $q^2$ bins ( $0 < q^2 < 12$ in GeV$^2$), and they are not consistent (at 1-$\sigma$) with the ones presented including all the lattice results. Note that these two types of predictions are consistent in the high-$q^2$ regions ($q^2 \gsim$ 12 GeV$^2$ ). We have noticed a difference in the shapes of the form factors between the two types of fits in the low $q^2$ regions (Fig.~\ref{fig:FFBsK}). Furthermore, the inclusion of $f_{+,0}^{B_s\to K}(q^2=0)$ from LCSR changes the values of the branching fractions in the lower bins and for the bin $[0,2]$ GeV$^2$ the predicted values are not consistent with each other at 1-$\sigma$. This difference is due to the inconsistency of the LCSR input with the lattice-predicted shape of the respective form factors at $q^2=0$ (Fig.~\ref{fig:FFBsK}). The predicted values are marginally consistent up to $q^2$ $\sim$ 12 GeV$^2$, and above that, they are consistent at 1 $\sigma$. Note the predictions based on the inputs from lattice and LCSR have 1-$\sigma$ errors of about $\lsim$ 10\% almost in all the bins. However, for the predictions with only lattice, the low $q^2$ bins have errors $\lsim$ 20\%. 
\begin{table}[t]
	\begin{center}
		\small
		\renewcommand*{\arraystretch}{1.6}
		\begin{tabular}{|c|c|c|c|}
			\hline
			\multicolumn{4}{|c|}{$\mathcal{BR}(\bar{B}_s \to K \ell^-\bar{\nu}_{\ell})\times 10^{5}$} \\
			\hline
			Bin[$\text{GeV}^2$] & With All Lattice  & All Lattice without $\text{HPQCD}^{\bar{B}_s \to K \ell^- \bar{\nu}}$  & With All Lattice + LCSR  \\
			\hline
			0 - 2 & 1.21 $\pm$ 0.22 & 0.46 $\pm$ 0.23 & 1.68 $\pm 0.21$\\
			2 - 4 & 1.25 $\pm$ 0.20 & 0.57 $\pm$ 0.23 & 1.67 $\pm$ 0.19\\
			4 - 6 & 1.27 $\pm$ 0.18 & 0.67 $\pm$ 0.21 & 1.62 $\pm$ 0.17\\
			6 - 8 & 1.28 $\pm$ 0.16 & 0.78 $\pm$ 0.19 & 1.56 $\pm$ 0.15\\
			8 - 10 & 1.28 $\pm$ 0.14 & 0.87 $\pm$ 0.17 & 1.51 $\pm$ 0.14\\
			10 - 12 & 1.27 $\pm$ 0.13 & 0.96 $\pm$ 0.15 & 1.44 $\pm$ 0.13\\
			12 - 14 & 1.24 $\pm$ 0.12 & 1.01 $\pm$ 0.13 & 1.36 $\pm$ 0.12\\
			14 - 16 & 1.18 $\pm$ 0.11 & 1.03 $\pm$ 0.11 & 1.26 $\pm$ 0.10\\
			16 - 18 & 1.07 $\pm$ 0.09 & 0.99 $\pm$ 0.09 & 1.12 $\pm$ 0.09\\
			18 - 20 & 0.90 $\pm$ 0.08 & 0.86 $\pm$ 0.08 & 0.92 $\pm$ 0.07\\
			20 - 22 & 0.60 $\pm$ 0.05 & 0.59 $\pm$ 0.05 & 0.62 $\pm$ 0.05\\
			22 - 23.75 & 0.17 $\pm$ 0.01 & 0.16 $\pm$ 0.01 & 0.17 $\pm$ 0.01\\
			\hline
		\end{tabular}
		\caption{Predictions of $\mathcal{BR} (\bar{B}_s \to K \ell^- \bar{\nu})$ in small bins of $q^2$, where $|V_{ub}|$ are taken from the fits to $\bar{B}\to \pi\ell^-\bar{\nu}_{\ell}$, i.e. from Fit-VIII or from Fit-IX in table \ref{tab:Vub} depending on the respective scenarios. These predictions are based only on the lattice or lattice and LCSR, respectively. }
		\label{tab:binpred1}
	\end{center}
\end{table} 

In table \ref{tab:binpred}, we have presented our predictions for the branching fractions $\mathcal{BR}(\bar{B}\to \rho(\omega) \ell^- \bar{\nu}_{\ell})$ in small $q^2$-bins. 
Here, for both the decay modes, the SM predictions are obtained using only the available LCSR inputs on the respective modes and $|V_{ub}|$ has been taken from the fit to $\bar{B}\to \pi\ell^- \bar{\nu}_{\ell}$ decays (experiment + lattice + LCSR). Currently, the SM predictions are based on only the LCSR inputs ($|V_{ub}|$ from $\bar{B}\to \pi$) and have large errors ($\lsim$ 20\%). Also, the measured values have large errors and are fully consistent with the predicted values based on LCSR. Predictions have also been obtained from the results of the combined fit to $\bar{B}\to (\pi,\rho,\omega)\ell^- \bar{\nu}_{\ell}$ decays, and these predictions have relatively small errors ($<$ 10\%). For $\bar{B} \to \rho\ell^- \bar{\nu}_{\ell}$ decays, these predictions from the results of the combined fit are lower than those obtained only from the LCSR inputs due to a lower value of $|V_{ub}|$ obtained in the combined fit. Note that for $\bar{B} \to \rho\ell^- \bar{\nu}_{\ell}$ decays, the predictions of the combined fit do not fully agree with the SM predictions based on LCSR and the respective measured values in all the bins. However, we have to keep in mind that the present measured values have large errors. In the absence of the possibility of NP in these modes, the predictions of the combined fits could also be considered as the SM predictions with relatively small errors. Apart from the very low $q^2$ bin, the SM predictions for $\mathcal{BR}(\bar{B} \to \omega \ell^- \bar{\nu}_{\ell})$ are consistent with the respective measured values. The predictions obtained from the combined fit result are also consistent with the respective SM predictions and the measured values except a few bins. The predictions obtained from the results of the combined fit have a lower value than the SM predictions only from the LCSR in this case as well due to the same reasons that have been discussed above.

	\begin{table} [t]
		\centering
		\small
		\renewcommand*{\arraystretch}{1.3}
		\begin{tabular}{|c|c|c|c|c|c|c|}
			\hline
			Bin[$\text{GeV}^2$] & \multicolumn{3}{|c|}{$\mathcal{BR}(\bar{B} \to \rho \ell^- \bar{\nu}$)$\times 10^{5}$ } & \multicolumn{3}{c|}{$\mathcal{BR}(\bar{B} \to \omega \ell^- \bar{\nu}$)$\times 10^{5}$}   \\
			\cline{2-7}
			& Experiment  & SM Prediction  & From fit results  & Experiment  & SM Prediction  & From fit results \\
			&  & (only LCSR) & & & (only LCSR) & \\
			\hline
			0 - 4 & 3.73 $\pm$ 1.10 & 5.28 $\pm$ 0.82 & 4.22 $\pm$ 0.32 & 1.38 $\pm$ 0.50 & 2.69 $\pm$ 0.69 & 1.65 $\pm$ 0.18\\
			4 - 8 & 7.18 $\pm$ 1.28 & 6.91 $\pm$ 1.08 & 5.44 $\pm$ 0.34 & 2.83 $\pm$ 0.46 & 3.50 $\pm$ 0.75 & 2.34 $\pm$ 0.16\\
			8 - 12 & 8.06 $\pm$ 1.37 & 8.06 $\pm$ 1.30 & 6.26 $\pm$ 0.37 & 2.85 $\pm$ 0.79 & 4.08 $\pm$ 0.82 & 2.87 $\pm$ 0.18\\
			12 - 16 & 7.23 $\pm$ 1.36 & 8.26 $\pm$ 1.43 & 6.32 $\pm$ 0.39 & 2.47 $\pm$ 0.94 & 4.18 $\pm$ 0.88 & 3.05 $\pm$ 0.25\\
			16 - 20 & 6.26 $\pm$ 1.24 & 5.85 $\pm$ 1.18 & 4.36 $\pm$ 0.34 & & &\\
			16 - 20.2 & & & & 2.40 $\pm$ 1.18 & 2.96 $\pm$ 0.75 & 2.21 $\pm$ 0.29\\
			20 - 20.285 & 0.17 $\pm$ 0.79 & 0.11 $\pm$ 0.03 & 0.08 $\pm$ 0.01 & & & \\
			\hline
		\end{tabular}
		\caption{Experimental data and SM predictions for $\bar{B} \to (\rho,\omega) \ell^- \bar{\nu}$ modes with $|V_{ub}|$ = (3.69 $\pm$ 0.12) $\times$ $10^{-3}$ (obtained from $\bar{B} \to \pi \ell^- \bar{\nu}$ modes). ``From fit results" corresponds to the fit results obtained from combined analysis with $\bar{B} \to (\pi,\rho,\omega) \ell^- \bar{\nu}$ modes.}
		\label{tab:binpred}
	\end{table}

\begin{table}[t]
	\begin{center}
		\small
		\begin{tabular}{|c|c|c|c|}
			\hline
			\multicolumn{4}{|c|}{Experimental : $R_{BF}$ (Low) = (1.66 $\pm$ 0.12) $\times$ $10^{-3}$ , $R_{BF}$ (High) = (3.25 $\pm$ 0.28) $\times$ $10^{-3}$ }\\
			\hline
			Mode & Inputs & \multicolumn{2}{c|} {Predictions of $R_{BF}$ ($10^{-3}$)} \\
			
			\cline{3-4}
			& & Low ($q^2$ $<$ 7 $\text{GeV}^2$) & High ($q^2$ $>$ 7 $\text{GeV}^2$) \\
			\hline
			$\bar{B} \to \pi l^- \bar{\nu}$  & Experiment+All Lattice  & 1.67 $\pm$ 0.11 & 3.32 $\pm$ 0.20 \\
			and		& Experiment+All Lattice+LCSR &  1.79 $\pm$ 0.10 & 3.17 $\pm$ 0.18\\
			\cline{2-4}
			$b \to c l^- \bar{\nu}$	& Experiment+All Lattice+LCSR  & 1.67 $\pm$ 0.11 & 3.30 $\pm$ 0.19 \\
			& (without $f_+^{B_s \to K}$) &   & \\
			\hline
			$b \to c l^- \bar{\nu}$, & Experiment+All Lattice+LCSR &  1.80 $\pm$ 0.10 & 3.08 $\pm$ 0.17 \\
			\cline{2-4}
			$b \to u l^- \bar{\nu}$	&  Experiment+All Lattice+LCSR  &  1.63 $\pm$ 0.11 & 3.30 $\pm$ 0.19 \\
			&  (without $f_+^{B_s \to K}$) &   & \\
			\hline
		\end{tabular}
		\caption{Estimated values of the ratio of the branching fractions $R_{BF}$ using the fit results of $|V_{ub}|/|V_{cb}|$ and the form factor parameters from the individual cases in table \ref{tab:ratio}.}
		\label{tab:pred}
	\end{center}
\end{table}
In table \ref{tab:pred}, we have presented our estimates of the ratio $R_{BF}$  corresponding to each fit scenario defined in table \ref{tab:ratio} using the respective fit results of the $|V_{ub}|/|V_{cb}|$ and the respective form factor parameters. The estimated values are consistent with the respective measured values. Also, here we note that the LCSR inputs $f_+^{B_s\to K}(q^2=0)$, do impact the ratios.

\bibliographystyle{JHEP}
\bibliography{ref_ASSI.bib}
\end{document}